\documentstyle[aps,twocolumn,floats,prd,psfig]{revtex}

\newcommand{\tot}{{\rm t}}
\def\C{{\bf C}}
\def\Ch{\widehat{\C}}
\newcommand{\hal}{{\rm h}}
\newcommand{\Poi}{{\rm P}}

\newcommand{\lin}{{}}
\def\sun{\hbox{$\odot$}}
\newcommand{\da}{d_A}
\newcommand{\wj}{\left(
                          \begin{array}{ccc}
                          l_1  &  l_2  & l_3 \\
                            0  &  0    &  0
                          \end{array}
                          \right)}
\newcommand{\wjm}{\left(
                          \begin{array}{ccc}
                          l_1  &  l_2  & l_3 \\
                           m_1  &  m_2   &  m_3
                          \end{array}
                          \right)}
\newcommand{\bi}{B_{l_1 l_2 l_3}}

\newcommand{\rad}{r}    
\newcommand{\sz}{{\rm SZ}}    

\begin{document}
\twocolumn[\hsize\textwidth\columnwidth\hsize\csname
@twocolumnfalse\endcsname

\title{Large Scale Pressure Fluctuations and Sunyaev-Zel'dovich Effect}
\author{Asantha Cooray}
\address{
Department of Astronomy and Astrophysics, University of Chicago,
Chicago, IL 60637. \\ E-mail: asante@hyde.uchicago.edu}

\date{Submitted to PRD -- May 2000}

\maketitle

\begin{abstract}
The Sunyaev-Zel'dovich (SZ) effect associated with pressure fluctuations of the
large scale structure gas distribution will be probed with current and
upcoming wide-field small angular scale cosmic microwave background 
experiments. We study the generation of
pressure fluctuations by baryons which are present in virialized dark
matter halos, with overdensities $\gtrsim 200$ 
and by baryons present in overdensities $\lesssim 10$. 
For collapsed halos, assuming the gas distribution is in hydrostatic equilibrium  with
matter density distribution, we predict the pressure power spectrum
and bispectrum associated with the large scale structure gas distribution by
extending the dark matter halo approach which describes the density
field in terms of correlations between and within halos.
The projected pressure power spectrum allows a determination of the
resulting SZ power spectrum due to virialized structures.
The unshocked photoionized baryons present in smaller
overdensities trace the Jeans-scale smoothed dark matter
distribution. They provide a lower limit to the SZ effect due to large scale structure in the absence of
massive collapsed halos. We extend our calculations to discuss higher order statistics, such as
bispectrum and skewness in SZ data. The SZ-weak lensing
cross-correlation is suggested as a probe of correlations between
dark matter and baryon density fields, while the
probability distribution functions of 
peak statistics of SZ halos in wide field CMB data can be used as a
probe of cosmology and non-Gaussian evolution of large scale 
pressure fluctuations.
\end{abstract}
\vskip 0.5truecm
]



\section{Introduction}

In recent years, increasing attention has been
given to the physical properties
of the intergalactic warm and hot plasma gas distribution
associated with large scale structure and the possibility of its
detection (e.g., \cite{CenOst99}).
It is now widely believed that at least $\sim$ 50\% of the
present day baryons, when compared to the total baryon density through
big bang nucleosynthesis, are present in this warm gas distribution
and have remained undetected given its nature (e.g., \cite{Fuketal98}). Currently proposed methods for the detection of this gas with
include observations of the
thermal diffuse X-ray emission (e.g., \cite{Pieetal00}), associated X-ray and UV absorption
and emission lines (e.g., \cite{Trietal00}) and resulting Sunyaev-Zel'dovich (SZ;
\cite{SunZel80}) effect (e.g., \cite{Cooetal00a}). 

The SZ effect arises from the  inverse-Compton scattering of CMB photons by hot electrons
along the line of sight. This effect has now been directly imaged
towards massive galaxy clusters (e.g., \cite{Caretal96,Jonetal93}), where temperature of the scattering medium  can reach as high as
10 keV producing temperature changes in the CMB of order 1 mK at
Rayleigh-Jeans wavelengths. Previous analytical predictions of the resulting SZ
effect due to large scale structure have been based on either through
a Press-Schechter (PS; \cite{PreSch74}) description of the
contributing galaxy clusters  (e.g., \cite{ColKai88,KomKit99})
or using a biased 
description of the pressure
power spectrum with respect to the dark matter 
density field (e.g., \cite{Cooetal00a}). Numerical simulations
(e.g., \cite{daS99,Refetal99,Seletal00}) 
are beginning to improve some of these analytical predictions,
but are still limited to handful of simulations with limited dynamical
range and resolution. Therefore, it is important that one consider
improving analytical models of the large scale structure SZ effect, and provide
predictions which can be easily tested through simulations.

Our present study on the large scale baryon pressure and the resulting
SZ effect is timely for several reasons, including the fact that improving numerical
simulations have recently begun to make detailed predictions for the pressure
power spectrum and SZ effect such that those predictions can be
extended and improved with analytical models
\cite{daS99,Refetal99,Seletal00}. Also, several studied have
considered the possibility that large scale baryon distribution can be
probed with upcoming CMB missions using SZ effect (e.g., \cite{Cooetal00a}).
Our calculations can be used to further refine these predictions and
to investigate the possibility how such analytical model as the one
presented here can be tested with observations.

As part of this study, we extend previous studies by considering the full power
spectrum and bispectrum, the Fourier space analog of the three-point
function, of pressure fluctuations. The pressure power spectrum and
bispectrum contains all necessary information on the large scale
distribution of temperature weighted baryons, whereas, the SZ power
spectrum is only a projected measurement of the pressure power
spectrum. This can be compared to weak gravitational lensing,
where lensing is a  direct probe of the projected dark matter density
distribution. The bispectrum of pressure fluctuations, and SZ
bispectrum, contains all the information present at the three-point
level, whereas conventional statistics, such as skewness, do not. 
An useful advantage of using the 3d statistics, such as the pressure
power spectrum, is that they can directly compared to numerical
simulations, while only 2d statistics, such as the projected pressure
power spectrum along the line of sight, basically the SZ power
spectrum, can be observed. Our approach here is to consider both
such that our calculations can eventually be compared to both
simulations and observations.

The calculation of pressure power spectrum and bispectrum requires
detailed knowledge on the baryon distribution, which can eventually be obtained
numerically through hydrodynamical simulations. Here, we provide an
analytical technique to obtain the pressure power spectrum and
bispectrum by describing the baryon distribution in the universe as
(1) present in virialized halos with overdensities $\gtrsim 200$ with
respect to background densities (2) unshocked diffuse baryons in
overdensities $\lesssim 10$ that trace a Jeans-smoothed dark matter
density field (3) the intermediate overdensity region, which is likely
to be currently undergoing in shock heating and falling on to
structures such as filaments.  
In the present paper we discuss the first two regimes, while a useful
approach to include the latter, through simulations, is discussed.

Our description of baryons present in virialized halos follow recent 
studies on the dark matter density field through halo contributions 
\cite{Sel00,MaFry00,Scoetal00} following \cite{SchBer91} 
and applied to lensing statistics in \cite{Cooetal00b} and
\cite{CooHu00b}. For the description of baryons, the critical
ingredients are: the PS formalism 
\cite{PreSch74} for the mass function; the NFW
profile of \cite{Navetal96}, and the halo bias
model of \cite{Moetal97}. The baryons are assumed to be in
hydrostatic equilibrium with respect to dark matter distribution, which is a
valid assumption, at least for the high mass halos that have been
observed with X-ray instruments, given the existence of regularity
relations between cluster baryon and dark matter physical properties
(e.g., \cite{MohEvr97}).
We take two descriptions of the temperature structure: (1)
virial temperature and (2) virial temperature plus 
an additional source
of minimum energy. The latter consideration allows the
possibility for a secondary source of energy for baryons, such as
due to preheating through stellar formation and feedback processes. 
Numerical simulations (e.g., \cite{CenOst99,Pen99}), as well observations (e.g., \cite{Davetal95,Ren97}), suggest the
existence of such an energy source. The low photoionized overdensity baryons are
described following the analytical description of \cite{GneHui98}. The fraction of baryons present in such low overdensities
are assumed to follow what has been measured in numerical simulations of
\cite{CenOst99}. We suggest that such baryons provide a lower
limit to the SZ effect in the absence of any contribution from baryons
present in virialized halos.

Throughout this paper, we will take $\Lambda$CDM as our fiducial cosmology 
with parameters $\Omega_c=0.30$ for the CDM density,
$\Omega_b=0.05$ for the baryon density, $\Omega_\Lambda=0.65$ for the
cosmological constant, $h=0.65$ for the dimensionless Hubble
constant and a scale invariant spectrum of
primordial fluctuations, normalized to  galaxy  cluster abundances 
($\sigma_8=0.9$ see \cite{ViaLid99})  
and consistent with COBE \cite{BunWhi97}.
For the linear power spectrum, we take the
fitting  formula for the transfer function given in \cite{EisHu99}.

The paper is organized as following: 
In \S \ref{sec:density}, we review the dark matter
halo approach to modeling the density field and extend it to model
properties associated with large scale baryon distribution, mainly the
pressure fluctuations that contributes to the observable SZ effect. 
We suggest recent papers by Seljak \cite{Sel00}, Ma \& Fry
\cite{MaFry00}, Cooray \& Hu \cite{CooHu00b}, and Scoccimarro et al
\cite{Scoetal00} for details on the dark matter halo approach
and applications to other observable statistics such as
galaxy properties and weak gravitational lensing.
As necessary, we use techniques developped in these papers for our
current calculation.
 In \S \ref{sec:sz}
we apply the formalism to the convergence power spectrum, skewness,
and bispectrum.  We conclude in \S \ref{sec:conclusions} with a
summary of our main results.

\begin{figure*}[t]
\centerline{\psfig{file=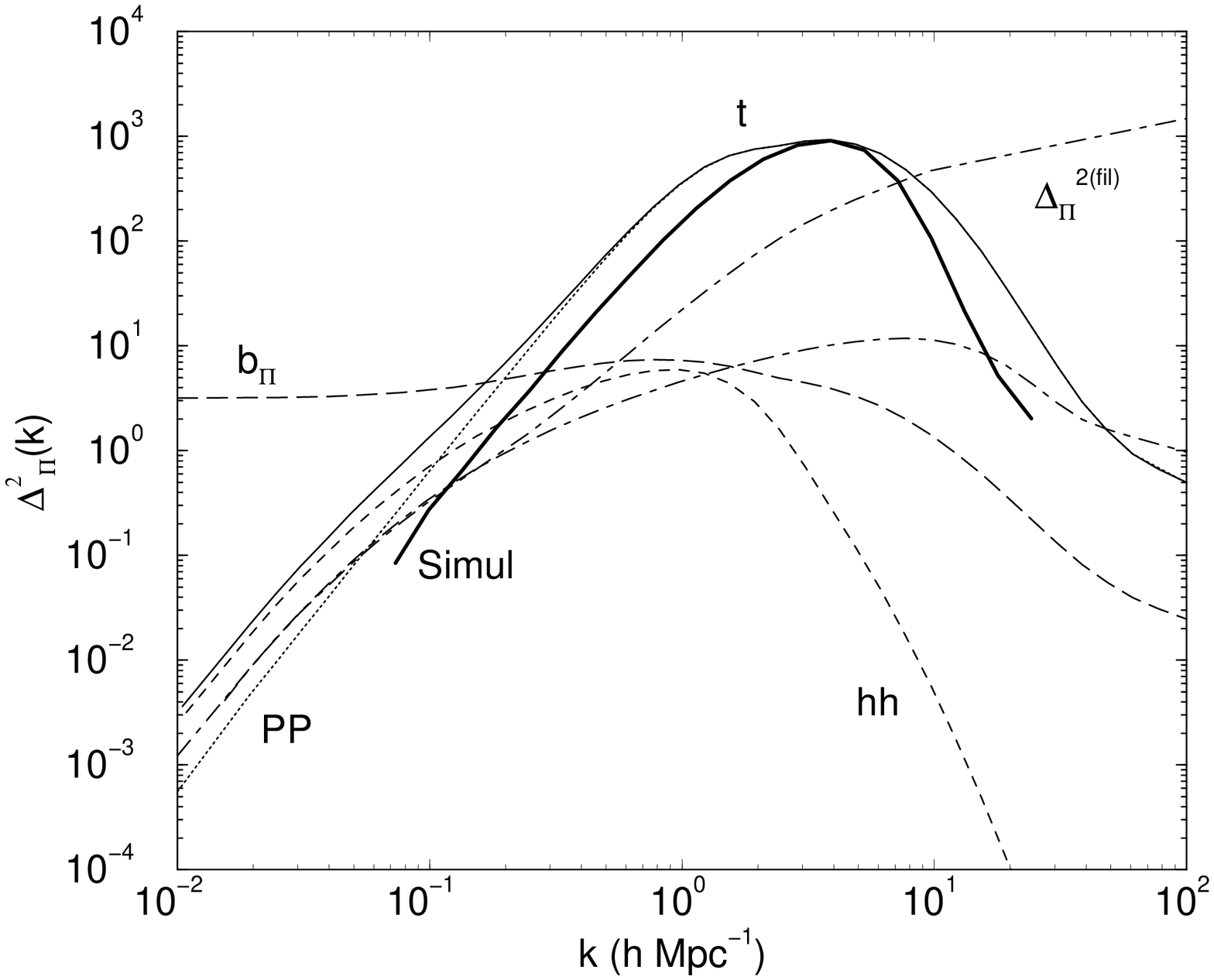,width=3.8in,angle=-90}
\psfig{file=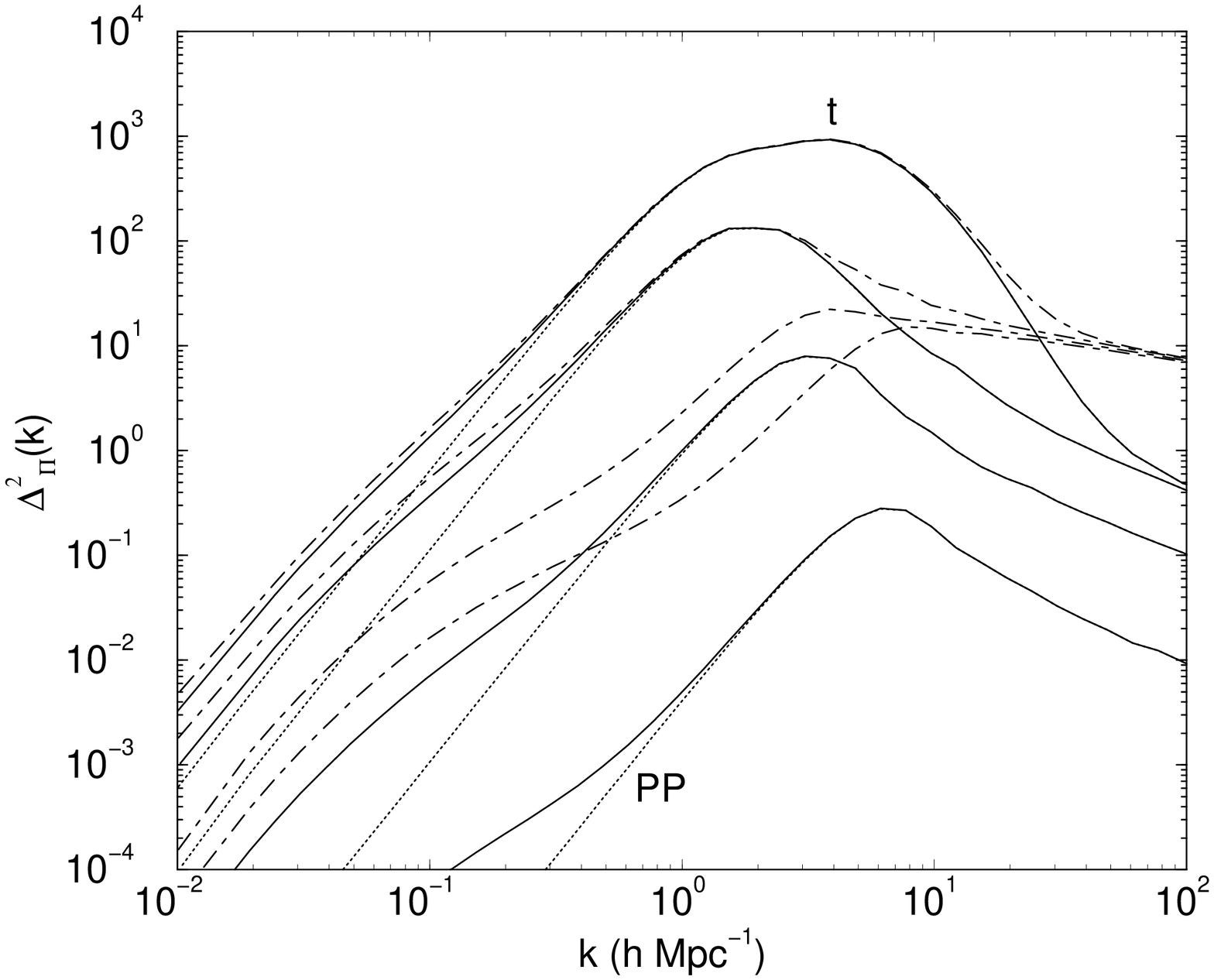,width=3.8in,angle=-90}}
\caption{Pressure power spectra with 
single halo (PP; dotted line), double halo (hh; dashed line)
and  total (t; solid line) contributions.
(a) Here, we use virial temperature to describe electrons and show
pressure bias, $b_\pi(k)$ (long dashed line),
filtered non-linear dark matter density power spectrum 
(\S~\ref{sec:gnehui})  and the
measured pressure power spectrum in simulations (thick solid line).  
(b) The variations in pressure power with maximum mass.
The dot-dashed lines 
show the total power with a minimum electron temperature of $0.75$
keV, as an attempt to reproduce power spectra under a possible preheating 
scenario.}
\label{fig:pressurepower}
\end{figure*}

\section{Density and Pressure Power Spectra}
\label{sec:density}
\subsection{General Definitions}

In order to calculate the contribution to temperature anisotropies
through SZ effect associated with large scale structure, we divide the
LSS with overdensities $\gtrsim 200$ as collapsed and virialized halos
with a gas distribution following hydrostatic equilibrium and with
virial temperatures. The pressure power spectrum can be calculated
using an extension to the  dark matter halo approach by assuming 
a physical relation between baryons and dark matter. 

The baryons with overdensities $\lesssim 10$
track the dark matter distribution and their power spectrum has been
studied by \cite{GneHui98}. These baryons have temperatures
similar to photoionization energies of Hydrogen and Helium.
Current numerical simulations, e.g., \cite{CenOst99}, 
suggest that most of the baryons are in such low overdensities at $z >
1$, while at present day, are in virialized halos.
The calculation of SZ effect due to such baryons follow
\cite{Cooetal00a}, except that we include the redshift
dependence of mass fraction within such low density halos following
the numerical results of \cite{CenOst99}, and modify the mean
temperature of such baryons to be consistent with photoionization energies.

First we discuss the pressure and related power spectra due to
collapsed halos.

The dark matter profile of collapsed halos are taken to be the NFW
\cite{Navetal96} with a density distribution
\begin{equation}
\rho_\delta(r) = \frac{\rho_s}{(r/r_s)(1+r/r_s)^{2}} \, .
\end{equation}
The density profile can be integrated and related to the total dark
matter mass of the halo within $r_v$
\begin{equation}
M_\delta =  4 \pi \rho_s r_s^3 \left[ \log(1+c) - \frac{c}{1+c}\right]
\label{eqn:deltamass}
\end{equation}
where the concentration, $c$, is $r_v/r_s$.
Choosing $r_v$ as the virial radius of the halo, spherical
collapse tells us that
$M = 4 \pi r_v^3 \Delta(z) \rho_b/3$, where $\Delta(z)$ is
the overdensity of collapse (see
e.g. \cite{Hen00}) and $\rho_b$ is the background matter density
today. We use
comoving coordinates throughout.
By equating these two expressions, one can
eliminate $\rho_s$ and describe the halo by its mass $M$ and
concentration $c$.

Following \cite{Cooetal00b}, we take the concentration
of dark matter halos to be
\begin{equation}
c(M,z) = a(z)\left[ \frac{M}{M_*(z)}\right]^{-b(z)}\,,
\label{eqn:concentration}
\end{equation}
where $a(z) = 10.3(1+z)^{-0.3}$ and $b(z) = 0.24(1+z)^{-0.3}$.
Here $M_*(z)$ is the non-linear mass scale at which the peak-height
threshold, $\nu(M,z)=1$.
The above concentration is chosen so that  dark matter halos
provide a reasonable match to the
the non-linear density power spectrum as predicted by the
\cite{PeaDod96};
it extends the treatment of \cite{Sel00} to the redshifts of
interest
for SZ effect.   We caution the reader that
eqn.~(\ref{eqn:concentration}) is only
a good fit for the $\Lambda$CDM model assumed.
 
The gas density profile, $\rho_g(r)$, is calculated assuming 
the hydrostatic equilibrium between the gas distribution and the dark
matter density field with in a halo. This is a valid assumption given
that current observations of halos, mainly galaxy clusters, suggest
the existence of regularity relations, such as size-temperature (e.g.,
\cite{MohEvr97}), between physical properties of dark matter and baryon distributions.
 
The hydrostatic equilibrium implies,
\begin{equation}
\frac{kT_e}{\mu m_p} \frac{d\log \rho_g}{dr} = -
\frac{GM_\delta(r)}{r^2} \, ,
\end{equation}
where now the $M_\delta(r)$ is the mass only out to a radius of $r$.
Note that we have assumed here an isothermal temperature for the gas
distribution.  Solving for the the equations above, we can
 analytically calculate the baryon density profile $\rho_g(r)$
\begin{equation}
\rho_g(r) = \rho_{g0} e^{-b} \left(1+\frac{r}{r_s}\right)^{br_s/r} \, ,
\label{eqn:gasprofile}
\end{equation}
where $b$ is a constant, for a given mass,
\begin{equation}
b = \frac{4 \pi G \mu m_p \rho_s r_s^3}{k_B T_e} \, ,
\label{eqn:b}
\end{equation}
with the Boltzmann constant, $k_B$. 

In general, the halos are described with virial temperatures
\begin{equation}
k_B T_e =  \frac{\gamma G \mu m_p M_\delta(r_v)}{3 r_v} \, ,
\label{eqn:virial}
\end{equation}
with $\gamma=3/2$ and $\mu=0.59$.
In addition, we also consider the possibility for the existence of a
constant non-gravitational energy in small mass halos, consistent with
observations of galaxy groups, due to what is commonly known as
``preheating''. The possibility for such a minimum energy for baryons
today comes from heating before virialization due to
energy injection and feedback processes such as processes associated
with
stellar formation.
The total gas mass present in a dark matter halo within $r_v$ is
\begin{equation}
M_g(r_v) = 4 \pi \rho_{g0} e^{-b} r_s^3 \int_0^{c} dx \, x^2
(1+x)^{b/x} \, .
\label{eqn:gasmass}
\end{equation}

The physical properties of the profile defined in Eq.~(\ref{eqn:gasprofile}) 
for baryons within dark matter halos, and a comparison to commonly
used profiles such as isothermal and so-called beta-profiles, can be
found in \cite{Maketal98} and \cite{Sutetal98}.
Compared to conventional profiles, this profile has the advantage that
it is directly related to the dark matter profile parameters, such as
central density $\rho_s$ and concentration via scale radius $r_s$,
thus, any changes to the dark matter distribution produces resulting
changes in the baryon distribution.
Also, one can study the effect of temperature variations on the gas distribution as the
parameter $b$ defined in Eq.~(\ref{eqn:b}) depends on it. A proper
normalization for the dark matter halo distribution containing baryons
comes through the $c(M,z)$ relation in
Eq.~(\ref{eqn:concentration}) such that the non-linear dark matter power
spectrum is produced in numerical simulations by same halos.
Note that our hydrostatic equilibrium ignores the self-gravity
contribution
from baryons to the total potential as we only include the dark matter
contribution to total mass. Since the baryon mass is expected to be
$\lesssim 10$\% of the total mass, 
we can safely ignore, as a first approximation, the
contribution to total mass from baryons themselves.

Roughly speaking, the perturbative aspect of the
clustering of the dark matter and baryons is described by the correlations between
halos, whereas the nonlinear aspect is described by the correlations
within halos, i.e. the halo profiles.
We will consider the Fourier analogies of the two and three point
correlations of the dark matter density, $\delta$, baryon pressure,
$\Pi$,  and galaxy distribution, $g$, defined in the usual way
\begin{equation}
\left< \delta_i^*({\bf k}) \delta_i({\bf k}') \right> = (2\pi)^3
\delta({\bf k}-{\bf k}') P_i^\tot(k) \, ,
\end{equation}
\begin{equation}
\nonumber
\left< \delta_i({\bf k}_1) \delta_i({\bf k}_2) \delta_i({\bf k}_3) \right> =
(2\pi)^3 \delta({\bf k}_1+{\bf k}_2 + {\bf k}_3) B_i^\tot(k_1,k_2,k_3)
\, ,
\end{equation}
with $i$ representing  $\delta$, $\Pi$ or $g$. We will also consider
cross-correlations
between the two, such as the dark matter density-pressure power
spectrum $P_{\delta\Pi}^\tot(k)$, which is what one probes by
correlating, say, the SZ effect and weak gravitational lensing
observations. Here and throughout, we occasionally suppress the redshift dependence
where no confusion will arise.

As presented in \cite{CooHu00b},
these spectra are related to the {\it linear} density power spectrum
$P(k)$ through
the bias parameters and the normalized 3d Fourier transform of the
density profile $\rho_i(r,M)$
\begin{equation}
y_i(k,M) = \frac{1}{M_i} \int_0^{r_v} dr\, 4 \pi r^2 \rho_i(r,M) \frac{\sin
(k
r)}{kr} \, ,
\end{equation}
where $i$ represents either the density, $\delta$, or the gas, $g$,
profile and associated masses respectively given in
equations~\ref{eqn:deltamass} and \ref{eqn:gasmass}. 
With an increase in temperature relative to the virial temperature of
the halo, especially for halos with masses $\lesssim 10^{13}$
M$_{\sun}$, the gas profile is such that it does not fall rapidly at the
virial radius, leading to an arbitrary cut off when doing the Fourier
transformation. We included an
additional filter to the gas density profile such that the gas density
profile decreases smoothly but promptly to zero at the virial radius:
$\rho'(r) = \rho(r) [{\rm erfc}(r-r_v/\sqrt{2}\Delta r)-1]$ with
$\Delta r \lesssim r_s$. Detailed aspects of the pressure and SZ statistics
due to medium to small mass halos ($\lesssim 10^{13}$ M$_{\sun}$) are
sensitive to sharpness of this transition, but these issues do not
change our primary results.  Here, we concentrate mostly on the
statistics
due to massive and rare halos. Another possibility not considered
here is to include the role of baryons at the outskirts of halos. 
The physical properties of such baryons can be semi-analytically
calculated following the assumption that virial radius provides a
shock boundary for the equilibrium of baryons within and outside
virialized regimes. The baryons outside halos are likely to preheated
and trace the Jeans-smoothed version of the dark matter density field.
The proper inclusion of such baryons require
the aid of numerical simulations or semi-analytical models.
These baryons are likely to include the ones present in
overdensities between 200 and 10, which we have neglected in the
present calculation.

Following \cite{CooHu00b}, it is convenient to define a
general integral over the halo mass function
$dn/dM$,
\begin{eqnarray}
&& I_{\mu,i_1\ldots i_\mu}^{\beta,\eta,\gamma}(k_1,\ldots,k_\mu;z) \equiv
\int dM \left(\frac{M}{\rho_b}\right)^\mu \frac{dn}{dM}(M,z)
b_\beta(M)  \nonumber\\
&& \left(\frac{\rho_b}{M}\frac{\left< N_g \right>}{\bar{n}_g}\right)^\gamma\times T_e(M,z)^\eta y_{i_1}(k_1,M)\ldots y_{i_\mu}(k_\mu,M)\,,
\label{eqn:I}
\end{eqnarray}
where $b_0 \equiv 1$.
\cite{Moetal97} gives the following analytic predictions for the
bias parameters
which agree well with simulations:
\begin{equation}
b_1(M;z) = 1 + \frac{\nu^2(M;z) - 1}{\delta_c} \, ,
\end{equation}
and
\begin{eqnarray}
b_2(M;z) &=& \frac{8}{21}[b_1(M;z)-1] + { \nu^2(M;z) -3 \over
\sigma^2(M;z)}\,.
\end{eqnarray}
Here, $T_e(M,z)$ is the electron temperature of the baryon
distribution of given halo when pressure power spectrum is considered,
$\nu(M,z) = \delta_c/\sigma(M,z)$, where
$\sigma(M,z)$ is the rms fluctuation within a top-hat filter at the
virial radius corresponding to mass $M$,
and $\delta_c$ is the threshold overdensity of spherical
collapse (see \cite{Hen00} for useful fitting functions).

The terms related to the galaxy power spectrum includes
the average number of galaxies per dark matter
halo, $\left< N_g \right>$, the mean number density of galaxies in the
universe $\bar{n}_g$. These parameters are discussed in
\S~\ref{sec:galgas} involving the galaxy-pressure power spectrum.

We use the Press-Schechter (PS; \cite{PreSch74})
mass function to describe $dn/dM$. We take the minimum mass
to be $10^3$ M$_{\sun}$ while the maximum mass is varied to
study the effect of massive halos on related statistics.
In general, masses above $10^{16}$ M$_{\sun}$ do not contribute to low
order statistics due to the exponential decrease in the number
density of such massive halos.

\begin{figure*}[t]
\centerline{\psfig{file=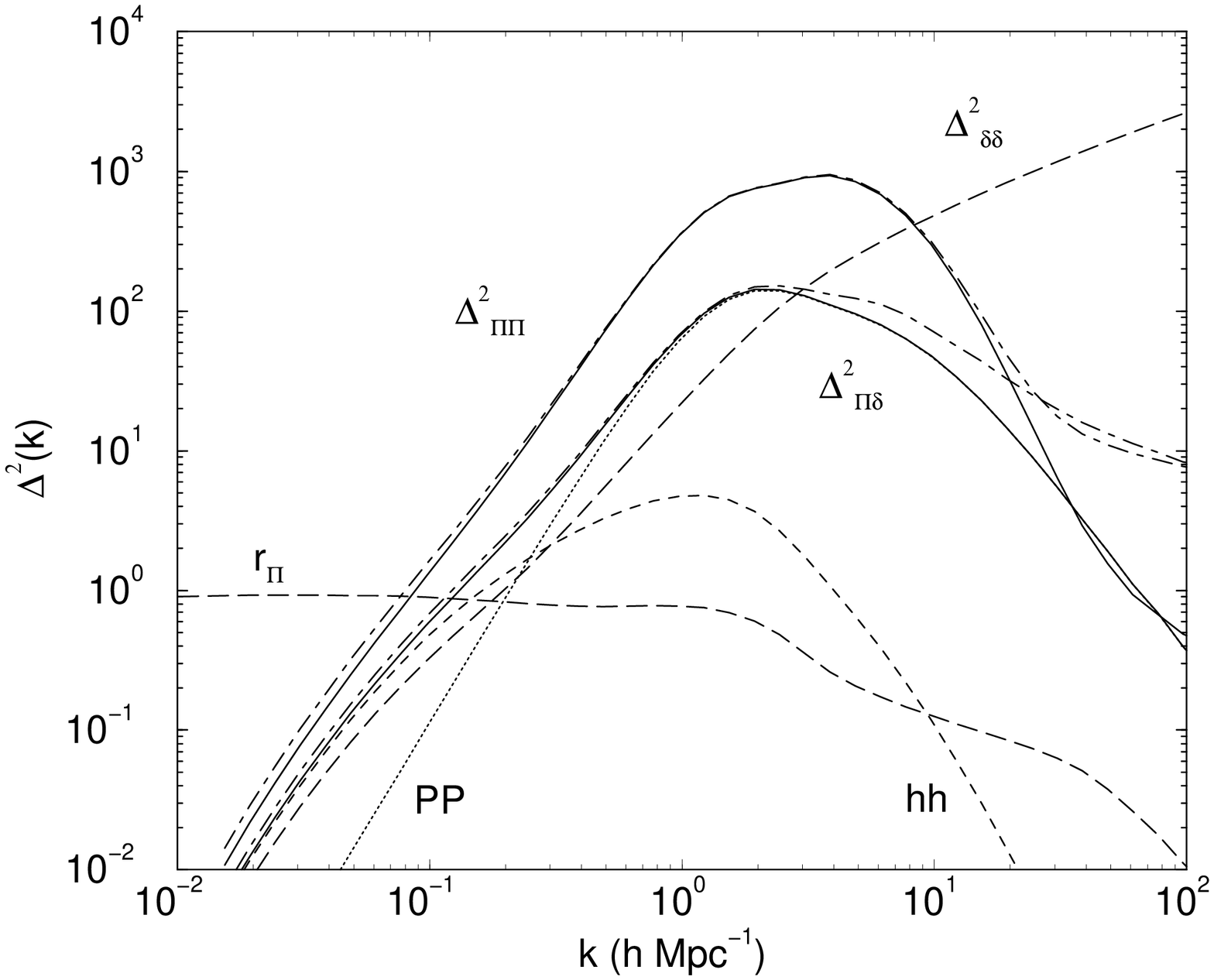,width=3.9in,angle=-90}
\psfig{file=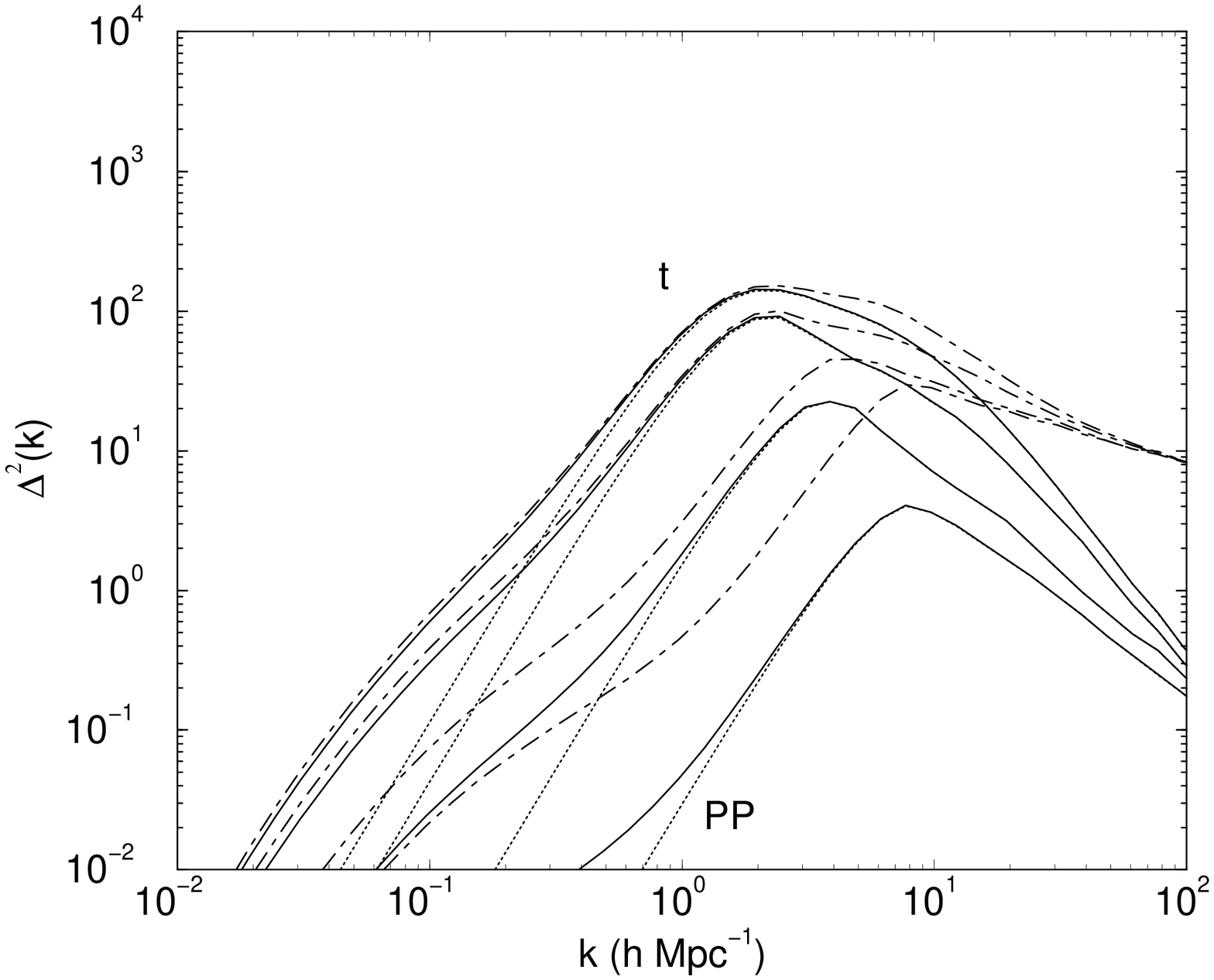,width=3.9in,angle=-90}}
\caption{Pressure-density cross-correlation power spectra. 
(a) Comparison of power spectra. $r_\Pi(k)$ is the correlation
between dark matter and baryon distributions.
(b) The variations in pressure power by changing the maximum mass
considered. The figure follows Fig.~\ref{fig:pressurepower}.}
\label{fig:pressuredmpower}
\end{figure*}

To summarize, in comparison to previous work on the SZ effect
from virialized halos, our model has following advantages: \\
(1) A physically motivated profile for the distribution of baryons in
virialized dark matter halos, instead of an assumed profile such as
the isothermal model or so-called $\beta$-profile.\\
(2) A mass function for the virialized structures with the dark matter
distribution of such halos reproducing the numerically simulated dark
matter power spectrum and higher order correlations.\\
(3) Easily extendable variations to the baryon physics so as to
account for issues such as pre-heating.\\
(4) Direct calculation of 3d properties of the large scale baryon
distribution, such as the pressure power spectrum and bispectrum,
which can be compared easily in numerical simulations.

We now discuss the calculation of properties related to the dark
matter and baryons. In the case of baryons, we discuss pressure
as this is the property that leads to the SZ effect allowing a useful
probe of them.

\subsection{Density Power Spectrum}
Following \cite{Sel00}, we can decompose the density
power spectrum, as a function of redshift, into contributions
from single halos (shot noise or ``Poisson'' contributions),
\begin{equation}
P_{\delta\delta}^{\Poi\Poi}(k) = I_{2,\delta\delta}^{0,0,0}(k,k) \,,
\end{equation}
 and correlations between two halos,
\begin{equation}
P_{\delta\delta}^{\hal\hal}(k) = \left[  I_{1\delta}^{1,0,0}(k) \right]^2 P^\lin(k)\,,
\end{equation}
such that
\begin{equation}
P_{\delta\delta}^\tot = P_{\delta\delta}^{\Poi\Poi} +  P_{\delta\delta}^{\hal\hal} \,.
\end{equation}
As $k \rightarrow 0$, $P_{\delta\delta}^{\hal\hal} \rightarrow P(k)$.

\subsection{Pressure Power Spectrum}

As above, we can decompose the pressure power
spectrum, as relevant for the SZ effect, into contributions from
single halos 
\begin{equation}
P_{\Pi\Pi}^{PP}(k) = I_{2,gg}^{0,2,0}(k,k) \,,
\end{equation}
and correlations between halos
\begin{equation}
P_{\Pi\Pi}^{hh}(k) = \left[  I_{1,g}^{1,1,0}(k) \right]^2 P^\lin(k)\,,
\end{equation}
such that
\begin{equation}
P_{\Pi\Pi}(k)^\tot = P_{\Pi\Pi}^{PP}(k) + P_{\Pi\Pi}^{hh}(k)
\end{equation}

For the pressure power spectrum, since $\eta$ in Eq.~(\ref{eqn:I}) is non-zero,
there is additional mass weighing arising from the fact that
$T_e \propto M^{2/3}$ resulting in an additional mass dependence.
The dependence is such that most of the contributions to the pressure,
and thus, to the SZ, power spectrum comes from most massive and rarest
halos. This dependence has already been observed in numerical
simulations by \cite{Seletal00}.

As we discuss later, this dependance on high mass halos to produce
most of the pressure fluctuations 
also leads to several interesting results with regards to
the detection and observability of SZ effect, among which are\\
(1) Most of the contribution to large scale SZ effect results
from massive clusters of galaxies, while smaller mass halos
and structures at low electron temperatures such as filaments do not
contribute significantly \\
(2) Since massive halos dominate the SZ effect, and the distribution
of these halos are Poisson and highly non-Gaussian, most of the
contribution to two point and higher-order statistics of
SZ effect will be dominated by Poisson term and there will be a
significant non-Gaussianity associated with large scale SZ effect\\
In fact, as we find later, the large scale correlations only
contribute at a level of 10\% to the SZ power spectrum suggesting that
such correlations can be mostly disregarded. The non-Gaussianity
associated with SZ effect may become a useful tool to separate out its
contribution from other sources of foregrounds in CMB anisotropy data,
though this task can be efficiently carried out using frequency
information (see, \cite{Cooetal00a}). We will return to all these
issues in later sections.

\begin{figure*}[t]
\centerline{\psfig{file=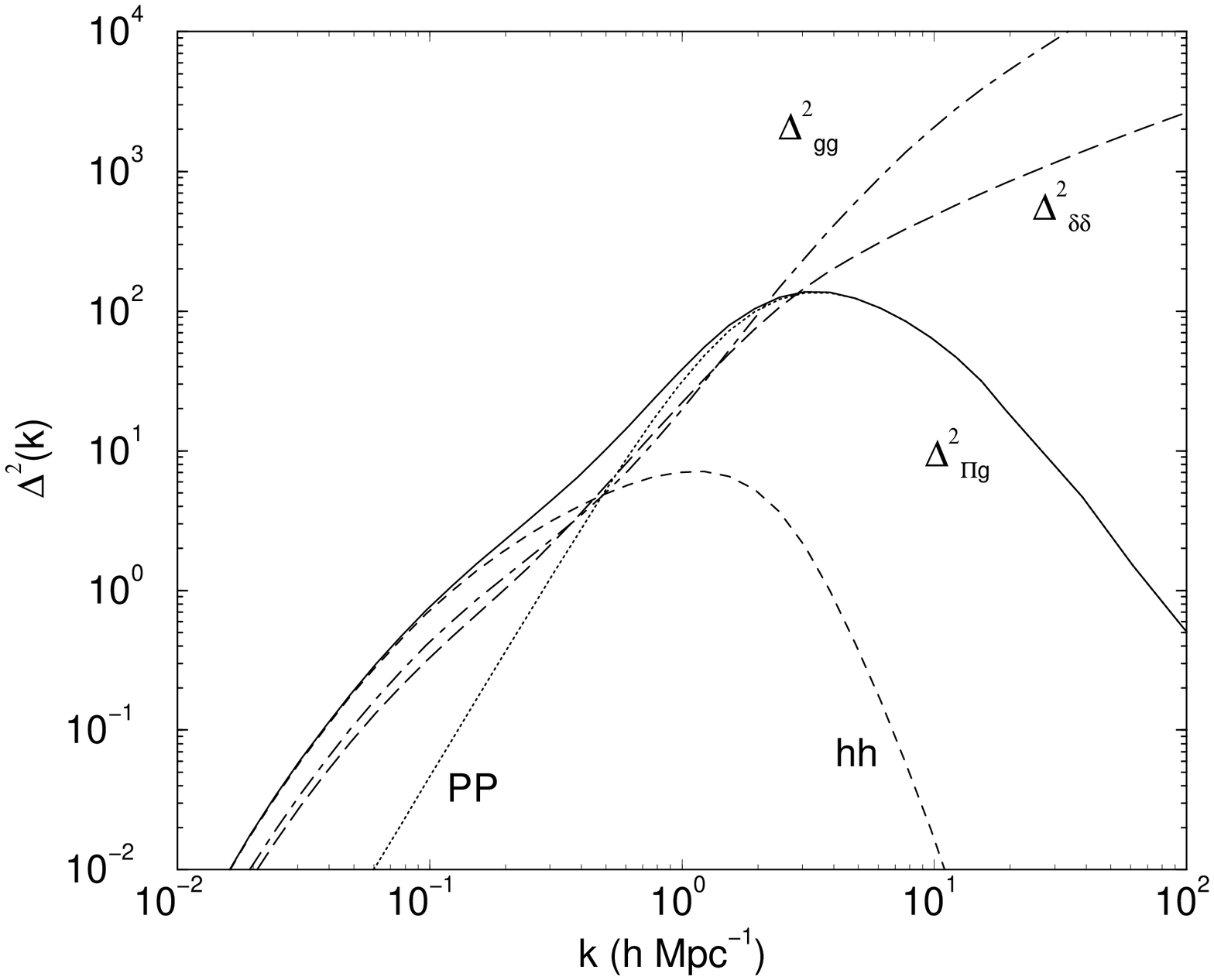,width=3.9in,angle=-90}
\psfig{file=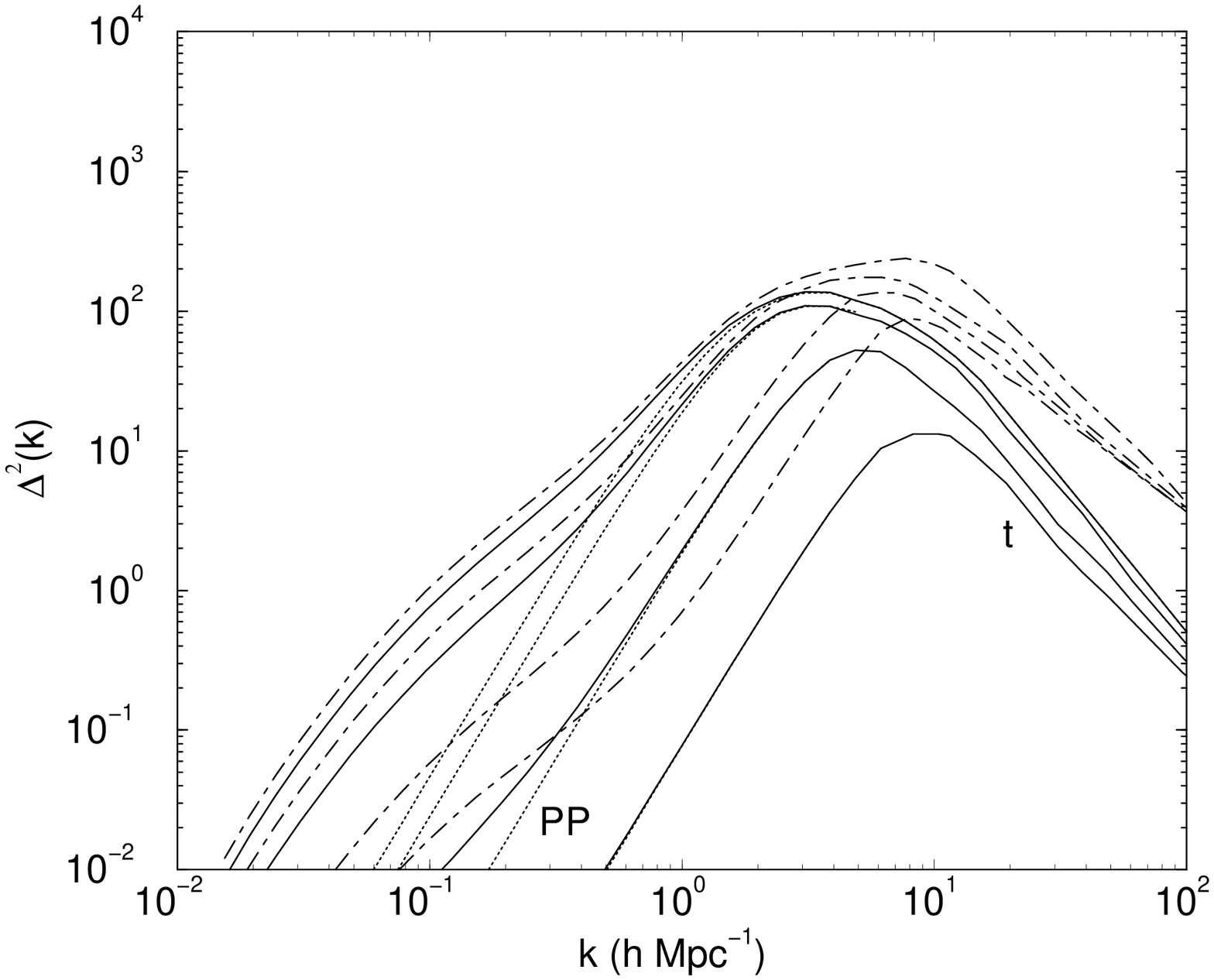,width=3.9in,angle=-90}}
\caption{Pressure-galaxy cross-correlation power spectra. 
(a) Comparison of pressure-gas power spectrum to the dark matter
density-density and galaxy-galaxy power spectra.
(b) The variations in pressure power by changing the maximum mass
considered. The figure follows Fig.~\ref{fig:pressurepower}.}
\label{fig:pressuregalpower}
\end{figure*}

\subsection{Density-Pressure Power Spectrum}

The cross-correlation between the density and gas field, as appropriate
for lensing-SZ cross-correlation can be decomposed as a single halo
\begin{equation}
P_{\Pi\delta}^{PP}(k) = I_{2,g\delta}^{0,1,0}(k,k) \,,
\end{equation}
and
\begin{eqnarray}
&&P_{\Pi\delta}^{hh}(k) = \left[  I_{1,g}^{1,1,0}(k) \right]
\left[ I_{1,\delta}^{1,0,0}(k)\right] P^\lin(k)\,,
\end{eqnarray}
such that
\begin{equation}
P_{\Pi\delta}(k)^\tot = P_{\Pi\delta}^{PP}(k) + P_{\Pi\delta}^{hh}(k)
\end{equation} 
 
With the pressure and density field power spectra, one can define
a bias associated with the large scale pressure, relative to density
field,
\begin{equation}
b_\Pi(k) \bar{T_e}  = \sqrt{\frac{P_\Pi(k)}{P_\delta(k)}} \, ,
\label{eqn:bias}
\end{equation}
and the dimensionless correlation coefficient between the dark matter
and baryon distributions
\begin{equation}
r_\Pi(k) = \frac{P_{\Pi\delta}(k)}{\sqrt{P_\delta(k)P_\Pi(k)}} \, .
\end{equation}

In Eq.~(\ref{eqn:bias}), the average density weighted temperature is
\begin{equation}
\bar{T_e}  = \int dM\, \frac{M}{\rho_p} \frac{dn}{dM}(M,z) T_e(M,z) \, .
\end{equation}

Following \cite{TegPee98}, one can define a covariance matrix
in Fourier space containing the full information on scale dependence
of bias and correlations:
\begin{equation}
\Ch(k )\equiv \left(\begin{array}{cc}
P_{\delta\delta}(k) & P_{\Pi\delta}(k) \\
P_{\Pi\delta}(k) & P_{\Pi\Pi}(k)
\end{array}\right) = P_{\delta\delta}(k)\left(\begin{array}{cc}
1 & b_\Pi r_\Pi \\ 
b_\Pi r_\Pi & b_\Pi^2
\end{array}\right) \, .
\end{equation}

The observation measurement of $b_\Pi$  and $r_\Pi$ can be considered
by an inversion of the SZ-SZ, lensing-lensing and SZ-lensing
power spectra as a function of redshift bins in which lensing-lensing
or SZ-lensing power spectra are constructed. We discuss these
possibilities later.

\subsection{Galaxy-Pressure Power Spectrum}
\label{sec:galgas}

The cross-correlation between the galaxy distribution and gas field, as appropriate
for galaxy-SZ cross-correlation can be decomposed as a single halo
\begin{equation}
P_{\Pi\delta}^{PP}(k) = I_{2,g\delta}^{0,1,1}(k,k) \,,
\end{equation}
and
\begin{eqnarray}
&&P_{\Pi\delta}^{hh}(k) = \left[  I_{1,g}^{1,1,0}(k) \right]
\left[ I_{1,\delta}^{1,0,1}(k)\right] P^\lin(k)\,,
\end{eqnarray}
such that
\begin{equation}
P_{\Pi\delta}(k)^\tot = P_{\Pi\delta}^{PP}(k) + P_{\Pi\delta}^{hh}(k)
\end{equation} 

The calculation of galaxy-pressure power spectrum requires knowledge
on the galaxy distribution within dark matter halos. Following,
\cite{Sel00}, we describe the average number of galaxies per
halo, $\left< N_g \right>$ in Eq.~(\ref{eqn:I}),  such that
\begin{equation}
\left< N_g \right> = \left\{ \begin{array}{cc}
  \left(\frac{M}{M_{\rm min}}\right)^{0.6} & {M \geq M_{\rm min}} \\
		 0  & {M < M_{\rm min}} 
 \end{array} \right.
\end{equation}
where $M_{\rm min}$, the minimum dark matter halo mass in which a
galaxy is found, is taken to be $5.3 \times 10^{11} h^{-1}$ M$_{\sun}$
for our fiducial $\Lambda$CDM cosmological model following
\cite{Benetal99}. The above relation is consistent with
semi-analytical models. however, we ignore scatter in the observed
distribution on the mean number of galaxies per halo.

With the the average number of galaxies per halo, as a function of
mass, the mean number density of galaxies can be written as an integral over
the PS mass function
\begin{equation}
\bar{n}_g = \int dM \, \left< N_g \right>\frac{dn}{dM}(M,z)
\, .
\end{equation}
In practice, the cross-correlation between galaxy distribution and any
other field requires the knowledge on the observable galaxy
properties such as the magnitude limit, relation between luminosity and
mass etc. The same restriction arising from observing conditions can
be introduced as part of the weight function that takes into account
the redshift projection of galaxies.

\begin{figure*}[t]
\centerline{\psfig{file=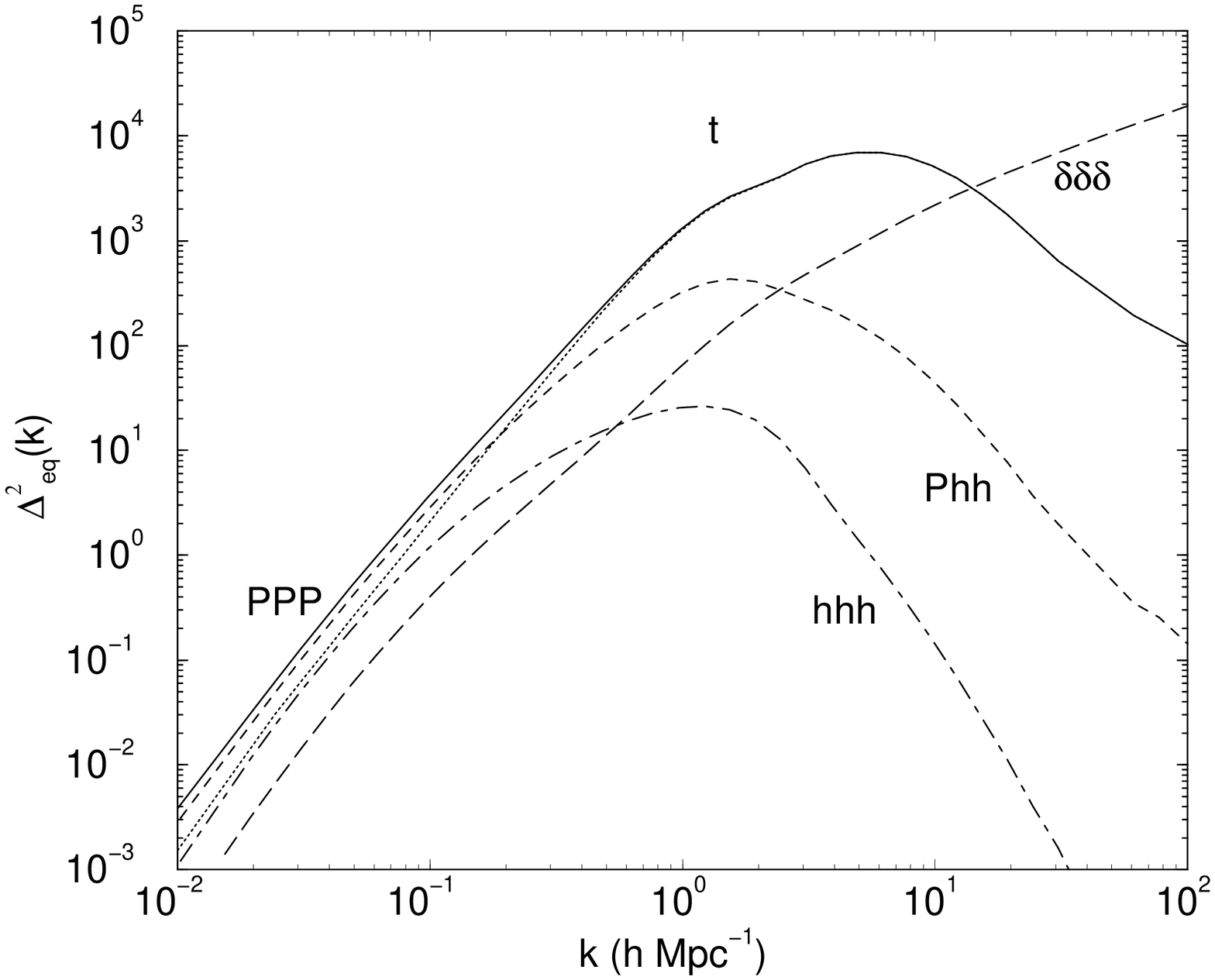,width=3.9in,angle=-90}
\psfig{file=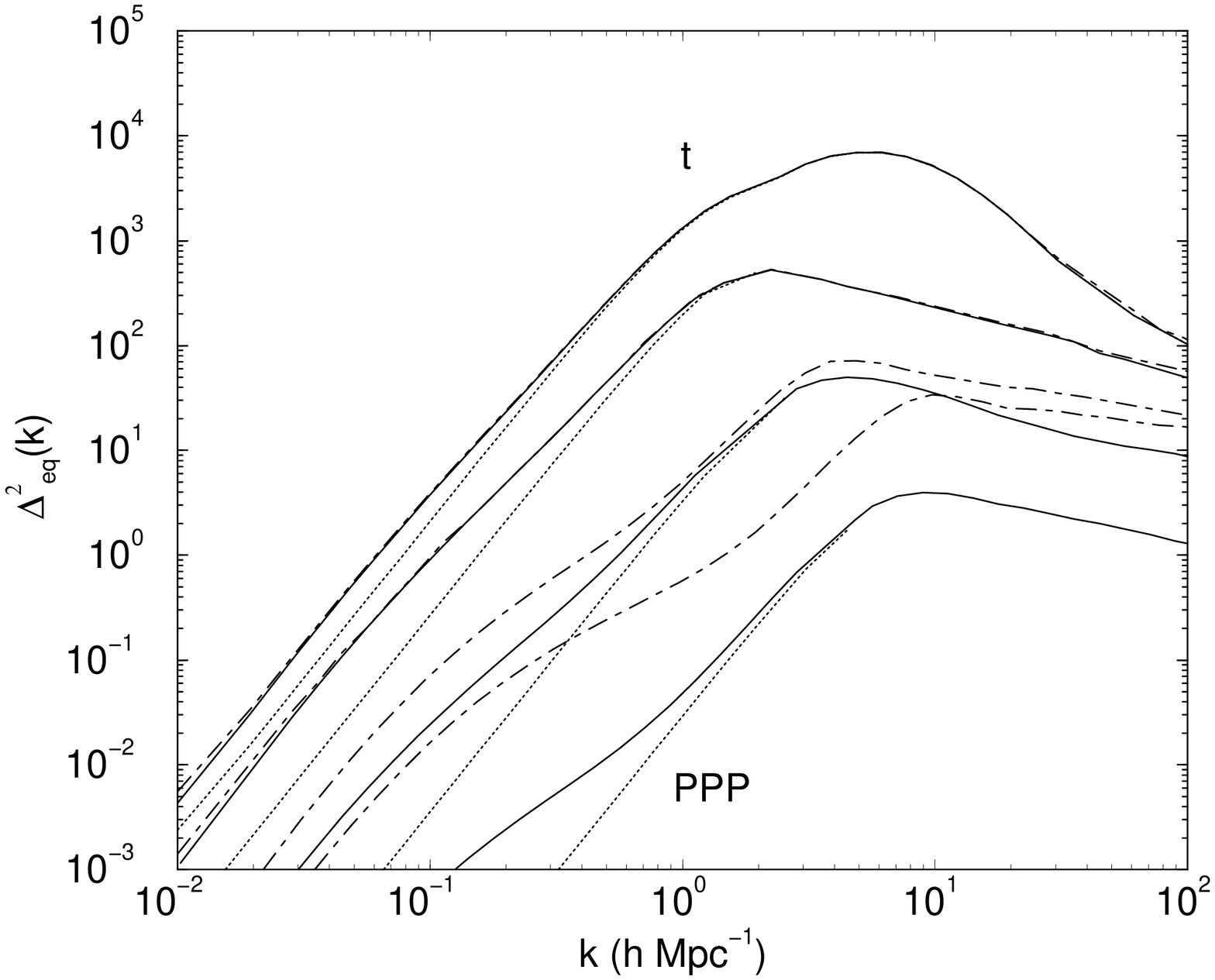,width=3.9in,angle=-90}}
\caption{Pressure bispectra. 
In (a), the long dashed line is the total
density field bispectrum. (b) The variations in pressure power by
changing the maximum mass considered. An increase in energy does not
significantly increase the contribution to the bispectrum when halos 
with high masses are considered. This is due to the
strong dependence on rare and most massive halos to the bispectrum
arising from single halo term.}
\label{fig:pressurebispectrum}
\end{figure*}

\subsection{Pressure Bispectrum}

Following \cite{CooHu00b}, we can write the
pressure bispectrum as
\begin{eqnarray}
B_\Pi^\tot &=& B_\Pi^{\Poi\Poi\Poi}  + B_\Pi^{\Poi\hal\hal}+  B_\Pi^{\hal\hal\hal} \,
,
\end{eqnarray}
where
\begin{eqnarray}
B_\Pi^{\Poi\Poi\Poi}(k_1,k_2,k_3) = I_{3,ggg}^{0,3,0}(k_1,k_2,k_3)\,,
\end{eqnarray}
for single halo contributions,
\begin{eqnarray}
B_\Pi^{\Poi\hal\hal}(k_1,k_2,k_3) = I_{2,gg}^{1,2,0}(k_1,k_2) I_{1,g}^{0,1,0}(k_3) P^\lin (k_3)
+
{\rm Perm.} \nonumber \\
\label{eqn:BPhh}
\end{eqnarray}
for double halo contributions, and
\begin{eqnarray}
B_\Pi^{\hal\hal\hal}(k_1,k_2,k_3) &=&
\left[ 2 J(k_1,k_2,k_3) I_{1,g}^{1,1,0}(k_3) + I_{1,g}^{2,1,0}(k_3) \right]
\label{eqn:Bhhh}
\\
&&\times
I_{1,g}^{1,1,0}(k_1) I_{1,g}^{1,1,0}1(k_2) P^\lin(k_1)P^\lin(k_2)
+ {\rm Perm.}
\nonumber
\end{eqnarray}
for triple halo contributions.
Here the 2 permutations are $k_3 \leftrightarrow k_1$, $k_2$.
Second order perturbation theory tells us that
\cite{Fry84,KamBuc99}
\begin{eqnarray}
J(k_1,k_2,k_3) &=& 1 - \frac{2}{7}\Omega_m^{-2/63}
                     + \left( { k_3^2 - k_1^2 - k_2^2 \over 2 k_1 k_2}
                     \right)^2
\nonumber\\
&&\times  \left[ \frac{k_1^2+k_2^2}{k_3^2 - k_1^2 -k_2^2}
+  \frac{2}{7}\Omega_m^{-2/63}\right] \, .
\end{eqnarray}

In addition to the pressure bispectrum, one can also define
cross-correlation bispectra such as the pressure-pressure-density or
pressure-density-density. These bispectra are relevant to the
calculation of non-Gaussianities present in CMB through secondary
anisotropies (e.g., \cite{CooHu00a}), and is necessary to
determine the higher order moments associated with cross-correlations
between individual effects such as SZ and weak lensing.

\subsection{Baryons in small overdensities}
\label{sec:gnehui}

The power spectrum of baryons that trace the Jeans-scale smoothed dark
matter density field can be calculated following \cite{GneHui98} (GH), where they studied  simple schemes to approximate the
effect of gas pressure. One such scheme that has fractional errors on the 10\% level for
overdensities $\lesssim 10$ is to filter the density perturbations in
the linear regime as $P_{b}^2 = f_b^2(k/k_{\rm F})
P_{\delta\delta}^2$ and treat the system as collisionless baryonic
particles.  Their best fit is obtained with the filter
\begin{equation}
f_b = {1 \over 2} [ e^{-x^2} + {1 \over (1+4 x^2)^{1/4}}]
\end{equation}
and \cite{GneHui98} suggests $k_{\rm F} = 34 \Omega_m^{1/2}
h$Mpc$^{-1}$
as a reasonable choice for the thermal history dependent filtering
scale.

For such baryons, we assume that their temperature is related to
photoionization energy ($\lesssim 25$ eV). The mass fraction of
baryons present in such small overdensities as a function of redshift
is obtained through the numerical simulations of \cite{CenOst99}.

\begin{figure*}[t]
\centerline{\psfig{file=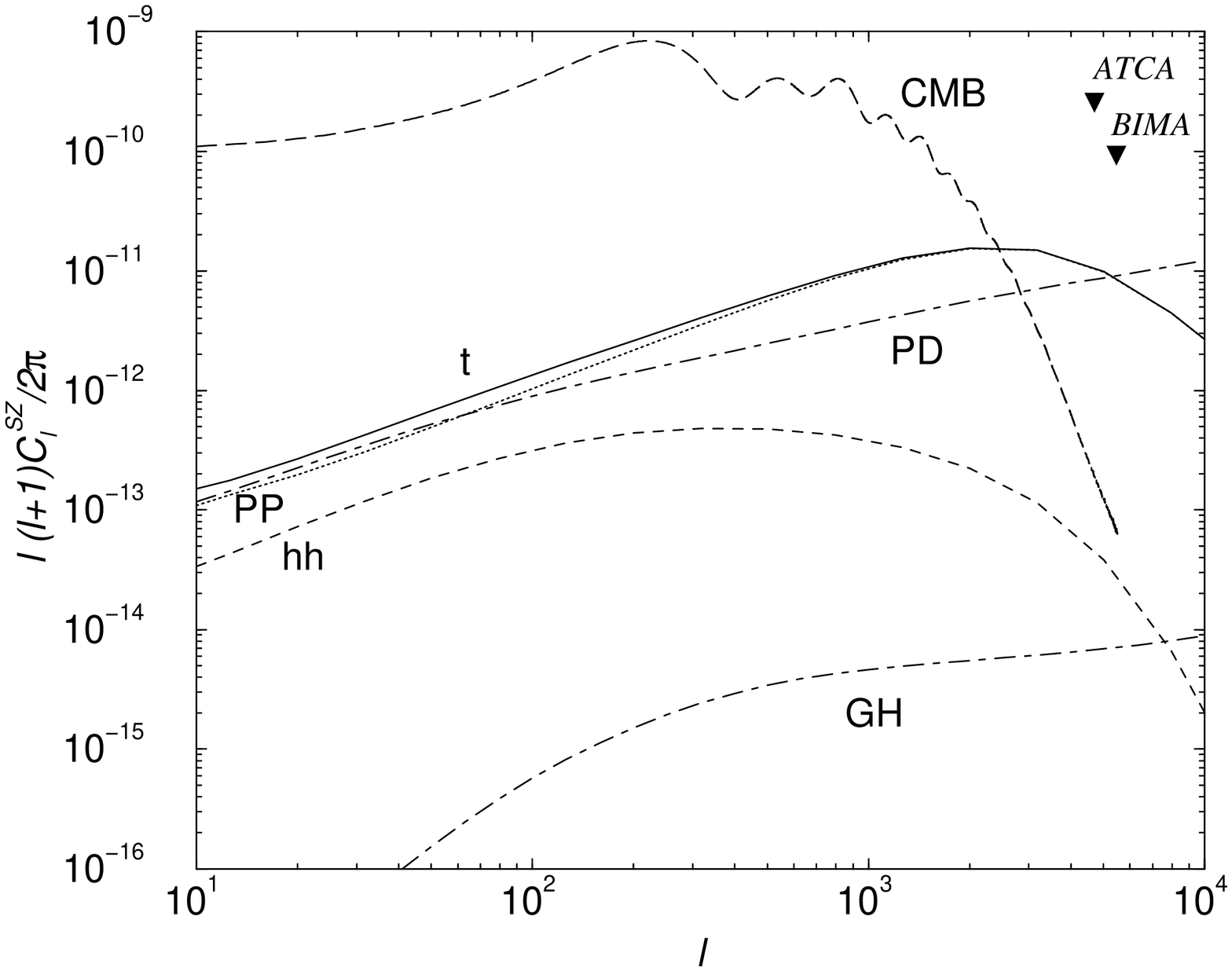,width=3.9in,angle=-90}
\psfig{file=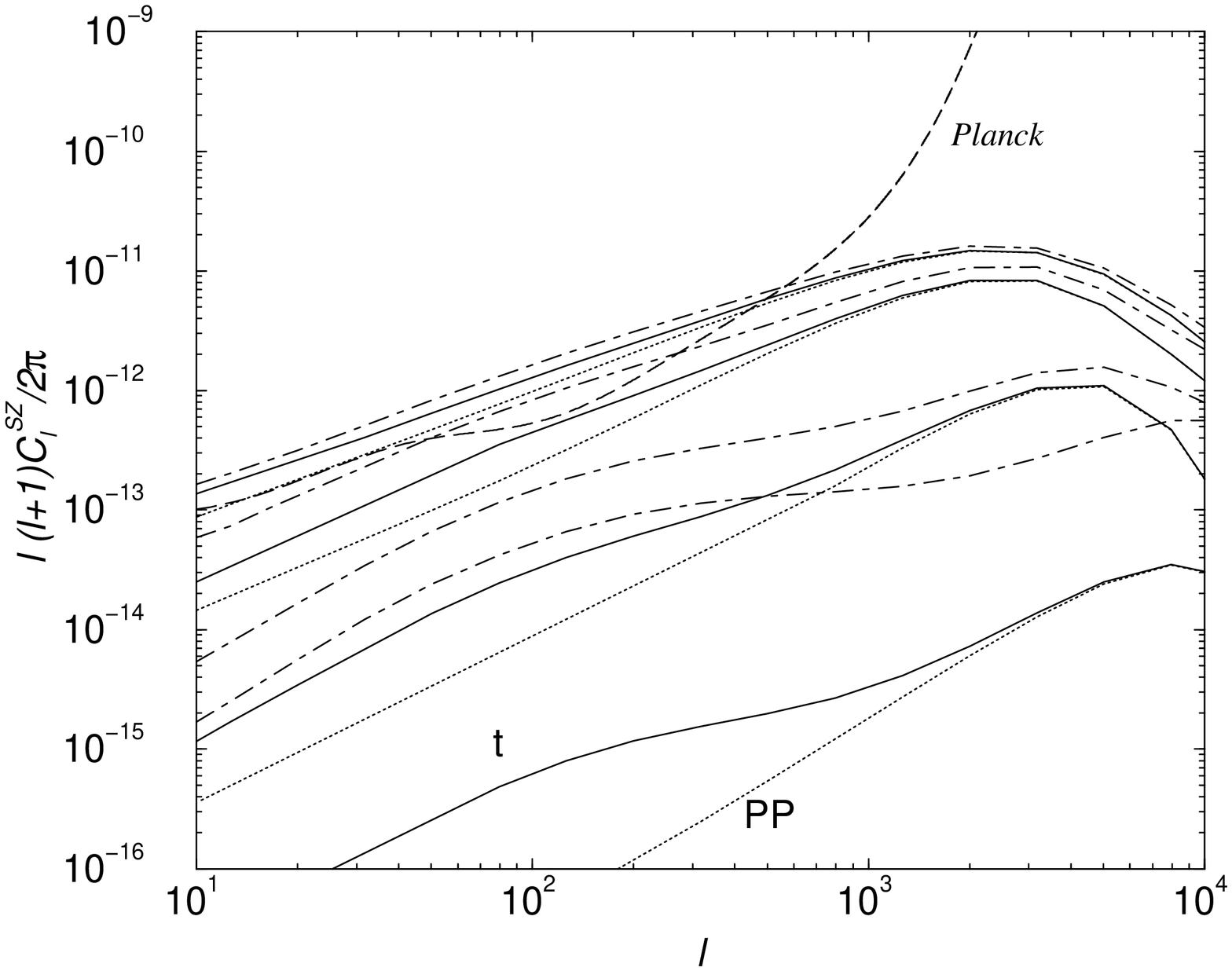,width=3.9in,angle=-90}}
\caption{SZ power spectrum. (a) The halo-halo correlations (dashed
line) contributes less than 10\% to the total SZ power (solid
line). The signal halo terms is shown with a dotted line. We also show
the SZ power spectra based on our previous prediction using  a
scale-independent bias model of PD non-linear dark matter power
spectrum and using the GH
Jeans-smoothed unbiased 
power spectrum to describe baryons present in small overdensities with
photoionization temperatures. For comparison, we also show the power
spectrum of lensing primary anisotropies and the upper limits on
temperature anisotropies at arcminute scales from BIMA and ATCA.
(b) The effect of maximum mass on the power
spectra, with maximum mass as in Fig.~1. The dot-dashed lines are the
total SZ power when the minimum temperature is 0.75 keV. We also show
the noise variance of the Planck SZ map, as has been calculated based
on SZ spectral dependance and Planck detector noise. The power
spectrum of SZ effect will be easily measured with Planck.}
\label{fig:szpower}
\end{figure*}

\subsection{Results \& Discussion}
\label{sec:results}

In Fig.~\ref{fig:pressurepower}(a-b), we show the pressure power power
spectrum today ($z=0$), written
such that $\Delta^2(k)=k^3 P(k)/2\pi^2$
is the power per logarithmic interval in
wavenumber.
In Fig~\ref{fig:pressurepower}(a), we show individual contributions
from the single and double halo terms
and a comparison to the Jeans-scale smoothed dark matter density field power
spectra, both linear and non-linear, following \cite{GneHui98} 
and using \cite{PeaDod96}  fitting function.
Here, we have taken the electron temperature to be the virial
temperature given in Eq.~(\ref{eqn:virial}).
Shown here is also the gas bias $b_\Pi(k)$; at large scales, $b_\Pi(k) \sim 3$
as $k \rightarrow 0$, consistent with numerically measured bias for
gas (e.g., \cite{Refetal99})  and analytical estimates in
\cite{Seletal00}. The pressure power spectrum is such that at
scales below $\sim$ few $h$ Mpc$^{-1}$, the pressure fluctuations are
suppressed relative to the dark matter power spectrum; The resulting
power spectrum can also be described as a smoothed, but biased,
version of the dark matter power spectrum. The scale at which
smoothing enters in to the power spectrum is determined by the
scale radius of the dark matter and gas profiles. Thus, the direct
measurement of the pressure power spectrum, to some extent, can be used as a probe of
halo profiles.  

In the same figure, we also show the measured pressure
power spectrum in hydrodynamical simulations by \cite{Refetal99}
 for their $\Lambda$CDM model. For comparative purposes, we 
have appropriately corrected
their pressure power spectra based on the mean temperature of baryons
as tabulated \cite{Refetal99} since our definition of the pressure power
spectrum includes the temperature. The resolution of simulations
limit the accuracy of power spectrum to the range in wavenumbers of
$0.2 \lesssim k \lesssim 2.0$ h Mpc$^{-1}$, and is only based on a
single realization. In this range, we find that
our analytical models predict more power than what is measured, 
while agreement is observed at scales of a $\sim$ few h Mpc$^{-1}$. 
The extent to which our analytical calculations agree with simulations
is encouraging; this is the first time that a detailed analytical
model for the pressure power spectrum has been compared with a
numerically measured one. Numerical simulations with improving dynamical range and resolution
will eventually test the reliability of models such as the one presented
here as a useful description of the pressure fluctuations of the
universe. Till then, we consider the present model as an appropriate
descrption of the large scale pressure fluctuations.

In Fig.~\ref{fig:pressurepower}(b), we show the dependence of pressure
power as a function of maximum mass used in the calculation with maximum
mass ranging from $10^{16}$, $10^{15}$, $10^{14}$
and $10^{13}$ M$_{\sun}$ from top to bottom.   Here, we
have shown the single halo contribution. Also shown are the total
contribution to pressure power spectrum when there is an additional
source of energy. Here, we have taken the minimum temperature to be
$\sim$ 0.75 keV; power spectra, in general, scale as the square of
this energy if the real preheating energy is higher or lower than the
one considered here. There are clear differences between the pressure
power spectra with and without an additional source of energy.
With increasing such additional non-gravitational energy, note that
$b$ in Eq.~(\ref{eqn:b}) $\rightarrow 0$ such that $\rho_g(r)
\rightarrow \rho_{g0}$. 
Thus, there is no longer a clear turn over in
the pressure power spectrum since the effect of smoothing resulting
from scale radius $r_s$ is not present.
The changes suggest the possibility that physical properties associated with
large scale structure baryons can be probed with pressure power spectrum.
In fact, the combined study of dark matter and pressure power
spectra may allow a consistent determination of halo properties, and to break
certain degeneracies associated with dark matter halo profile and
concentration as noted in \cite{Sel00}, while at the same time
investigating presence of additional sources of energy.

In Fig.~\ref{fig:pressuredmpower} and \ref{fig:pressuregalpower}, 
we study the cross-correlation power
spectra between pressure and density field and pressure and galaxy
distribution, respectively. These power spectra are relevant to the study
of correlations present between, say,  SZ and weak gravitational
lensing and SZ and galaxies, or a similar tracer of large scale
structure, such as radio sources. 
The presence of additional source of energy clearly affects
the cross-correlation power spectra, suggesting the possibility that
such effects may be investigated using cross-correlations between a
tracer of pressure fluctuations and a tracer of matter density fluctuations.

Since the bispectrum generally scales as the square of the power
spectrum,
it is useful to define
\begin{equation}
\Delta_{\rm eq}^2(k) \equiv \frac{k^3}{2\pi^2} \sqrt{B(k,k,k)} \,,
\end{equation}
which represents equilateral triangle configurations.
In Fig.~\ref{fig:pressurebispectrum}, we show the pressure bispectrum
as produced by baryons present in virialized halos. Here,  most of
the contributions at relevant  scales  come from the single halo
term. Given the additional dependence on temperature, and thus mass,
the bispectrum is more strongly sensitive to the presence of rare and
most massive halos. Thus, an increase in energy of such rare halos
does not significantly change the pressure bispectrum, but such energy
changes contribute when halos of mass $\lesssim 10^{14}$ M$_{\sun}$.

\begin{figure*}[t]
\centerline{\psfig{file=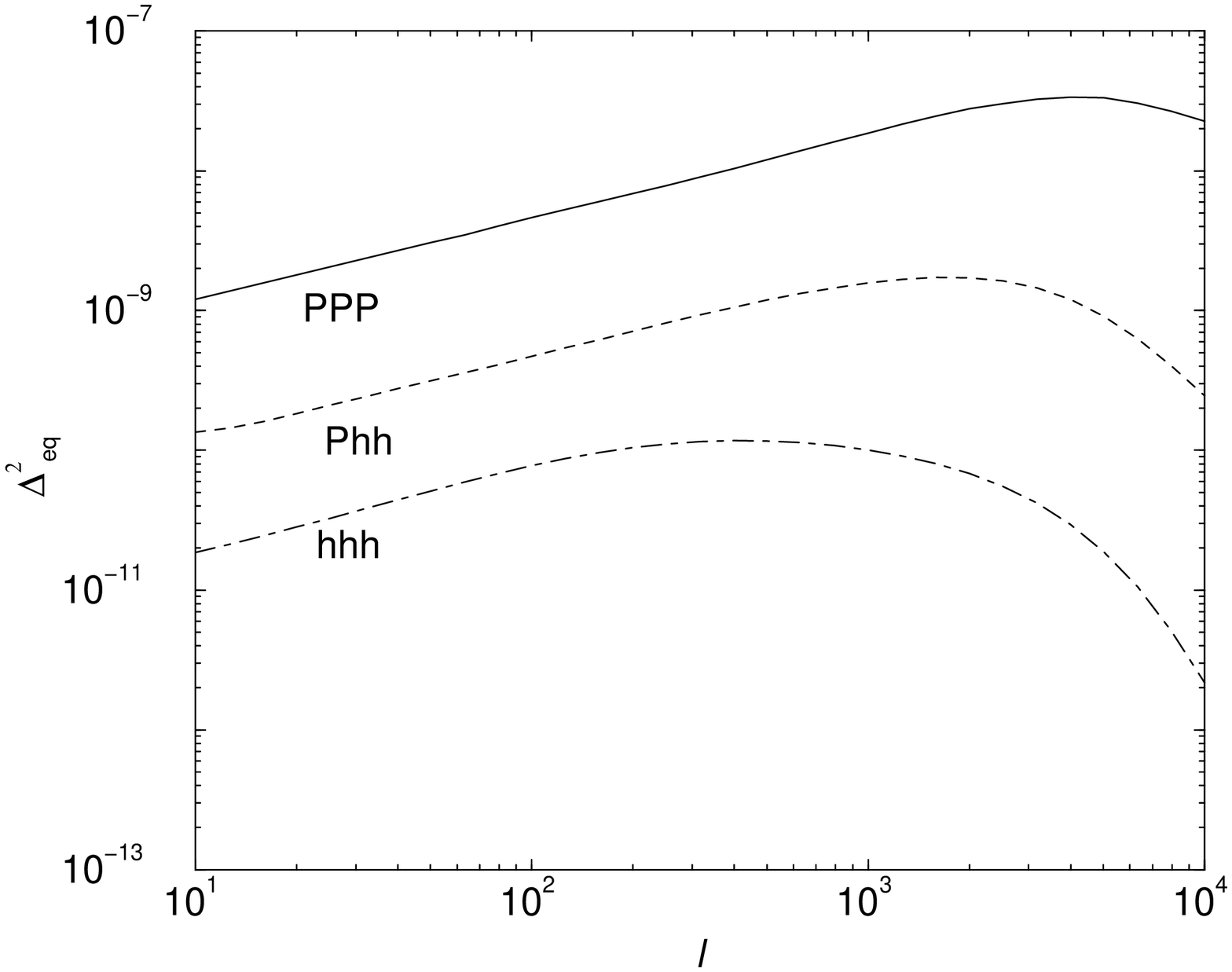,width=3.9in,angle=-90}
\psfig{file=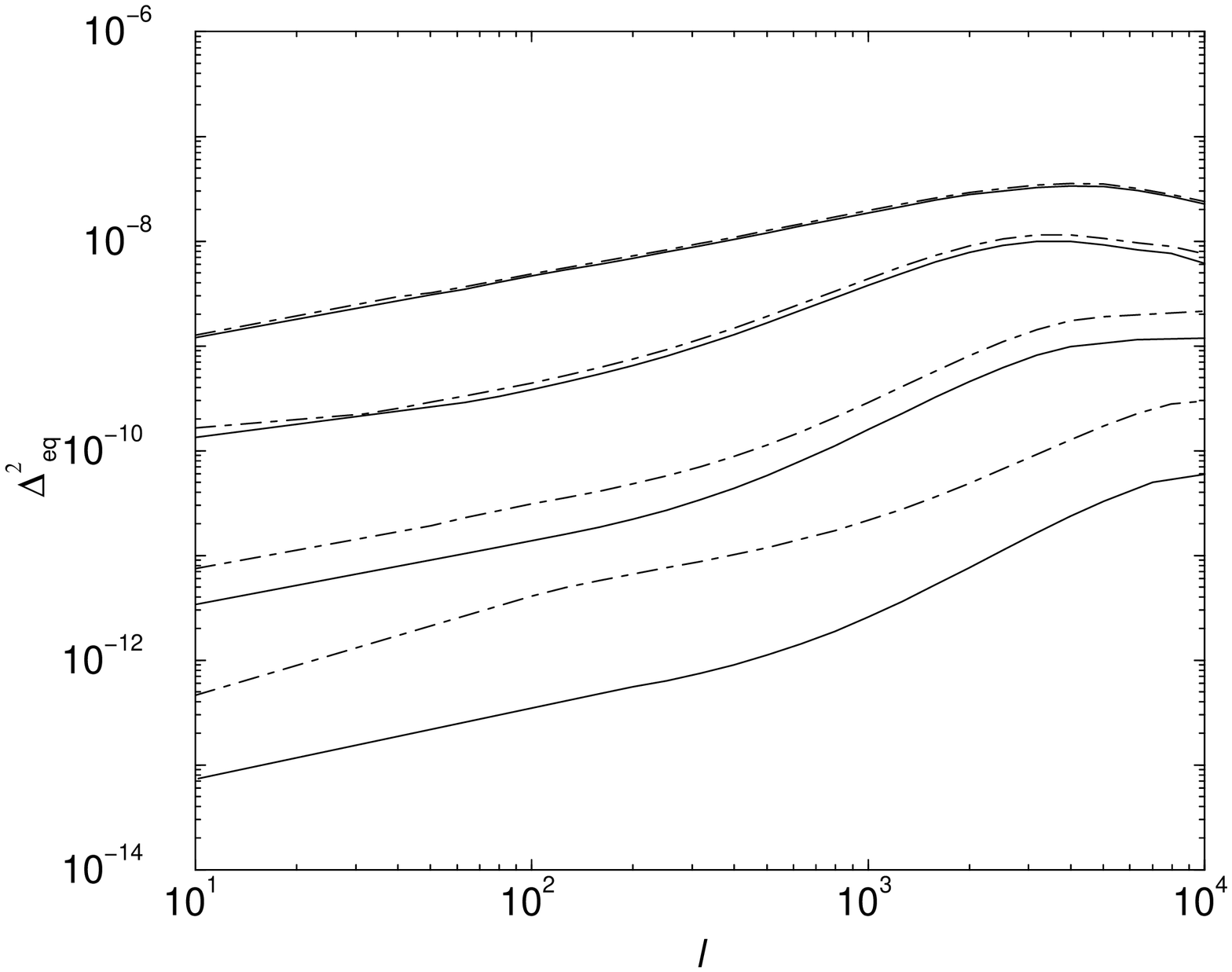,width=3.9in,angle=-90}}
\caption{SZ bispectrum. (a) The triple halo, halo-halo-halo, (dot dashed
line) and double halo, Poisson-halo-halo, correlations contributes
less than 10\% to the total SZ bispectrum (solid line). (b) The effect of
maximum mass on the bispectrum,  with maximum mass as in Fig.~1. The dot-dashed lines are the
total SZ bispectrum when the minimum temperature is 0.75 keV. Given
the strong dependence on mass here, the increase temperature does not
significantly change the bispectrum when maximum mass is greater than
10$^{15}$ M$_{\sun}$.}
\label{fig:szbispectrum}
\end{figure*}

\begin{figure*}[t]
\centerline{\psfig{file=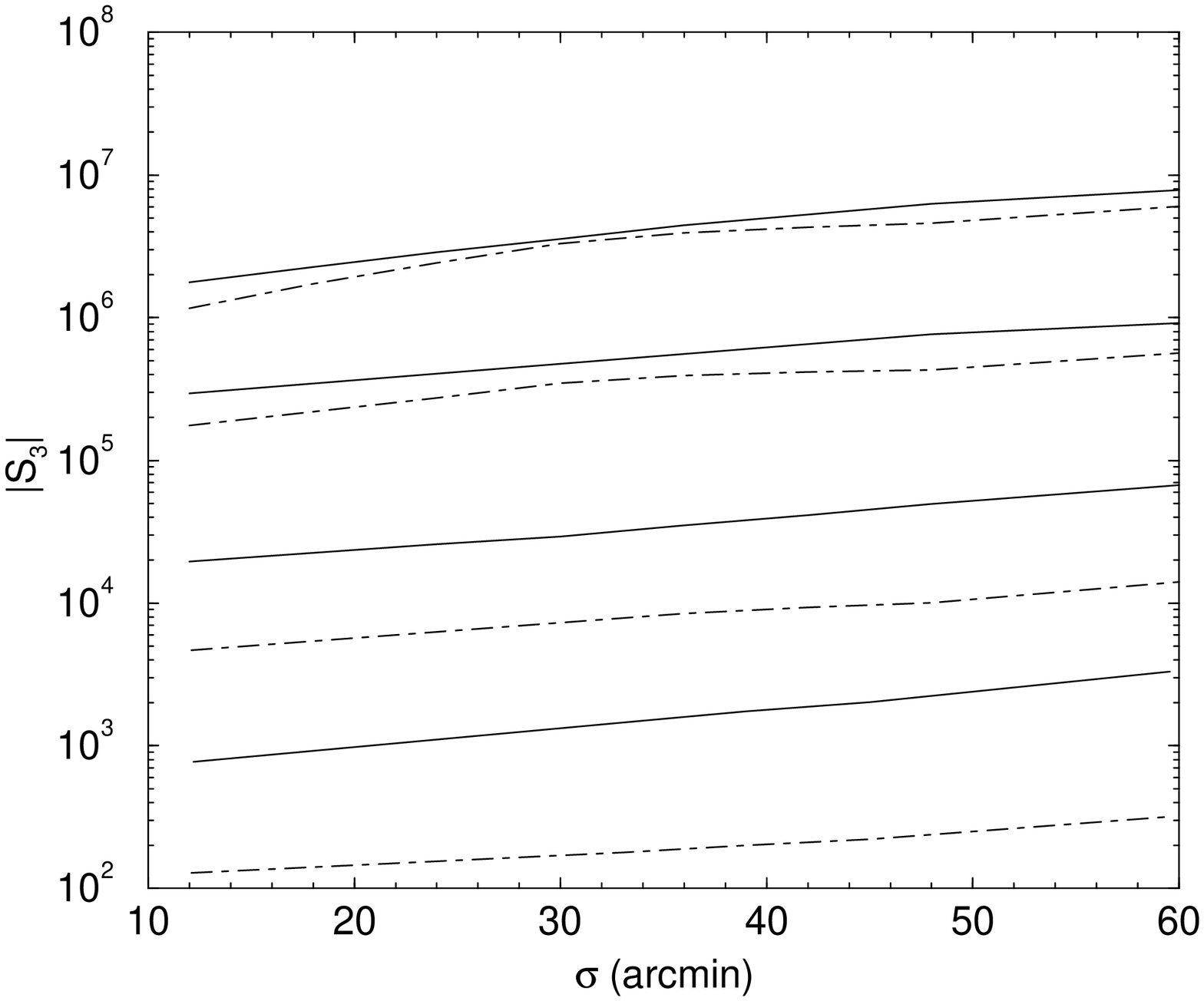,width=3.8in,angle=-90}
\psfig{file=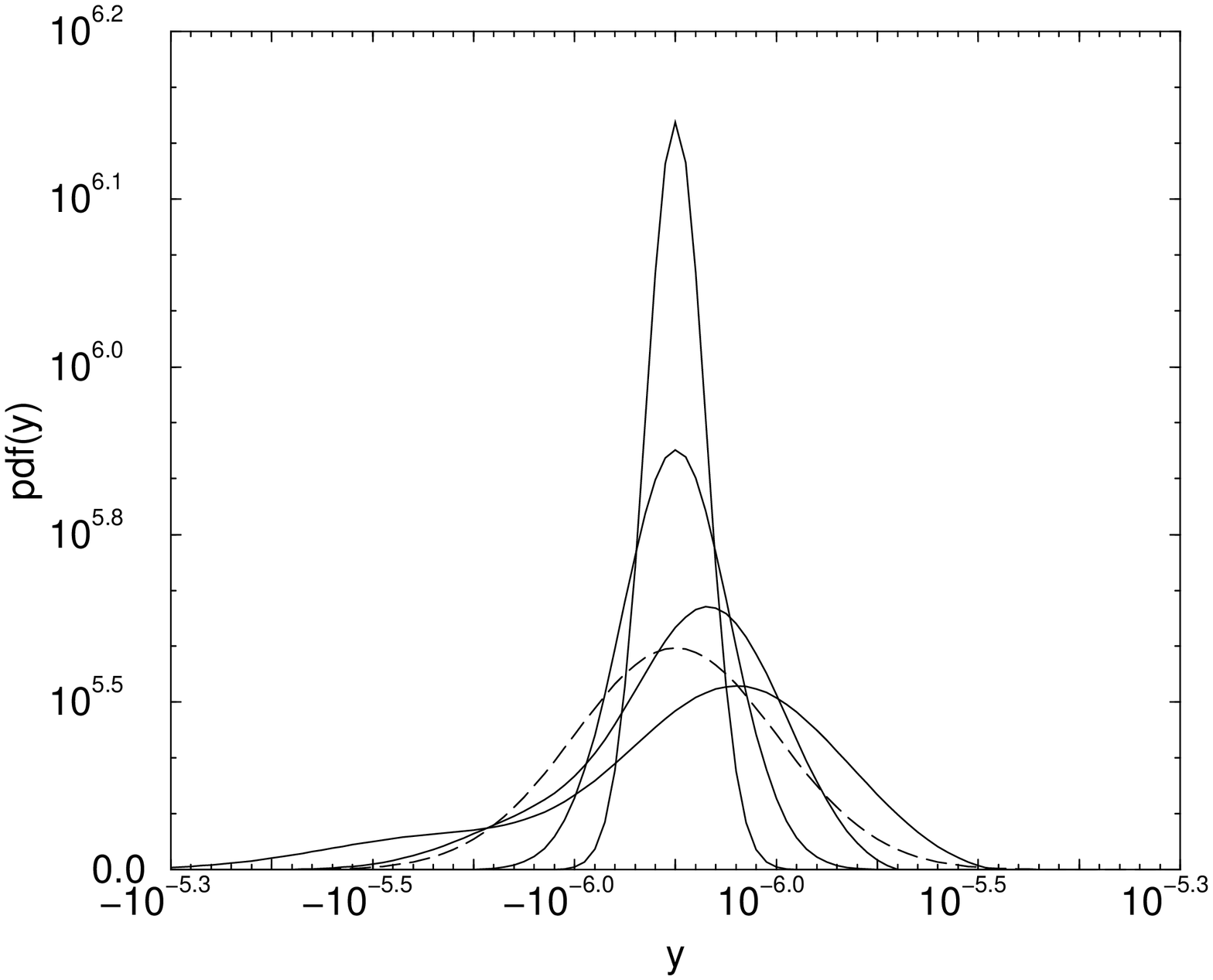,width=3.8in,angle=-90}}
\caption{(a) SZ skewness as a function of smoothing scale. The
absolute value of skewness, since $S_3 \propto g(x)^{-1} < 0$ as
$g(x)=-2$ at RJ part of the frequency spectrum,
is shown for the virial temperature (solid lines) and minimum
temperature (dot-dashed lines)  models for halos as a function of
maximum mass used in the calculation ranging from $10^{16}$ to
$10^{13}$ M$_{\sun}$ from top to bottom. (b) The probability distribution
function when the smoothing scale is 12 arcmin as a function of
maximum mass used (solid lines; the highest peak curve is when the
maximum mass is 10$^{13}$ M$_{\sun}$, while with increasing
non-Gaussianity as demonstrated by the departure from a Gaussian
distribution, the maximum mass increases). We also
show the pdf of expected Planck SZ map noise (dashed line)
for smoothing at same angular scales. The
non-Gaussian tail, at the negative $y$ values beyond the pdf of Planck
noise, due to massive and rare clusters will easily be
detected with Planck.}
\label{fig:szskewness}
\end{figure*}

\section{SZ effect}
\label{sec:sz}

The temperature decrement along the line of sight  due to SZ effect
can be written as the integral of pressure along the same line of sight
\begin{equation}
y\equiv\frac{\Delta T}{T_{\rm CMB}} = g(x) \int  d\rad  a(\rad) \frac{k_B
\sigma_T}{m_e c^2} n_e(\rad) T_e(\rad) \,
\end{equation}
where $\sigma_T$ is the Thomson cross-section, $n_e$ is the electron
number density, $\rad$ is the comoving distance, and $g(x)=x{\rm coth}(x/2) -4$ with $x=h \nu/k_B
T_{\rm CMB}$ is the spectral shape of SZ effect. At Rayleigh-Jeans (RJ)
part of the CMB, $g(x)=-2$.

The spectral dependance of SZ effect is unique in that it can be
separated from most other contributors to CMB temperature
fluctuations, including the primary anisotropy itself.
As discussed in Cooray et al. \cite{Cooetal00a}, the upcoming
multifrequency CMB satellite and ballone-borne data, among which
Planck providing the greatest information on SZ,
allow the possibility for detailed studies on the SZ effect including
its higher order correlations such as bispectrum and skewness.
Since these observations are projected measurements of the pressure
power spectrum and bispectrum along the line of sight, we now provide
analytical predictions based for SZ effect based on our model for the
pressure fluctuations.

\subsection{SZ Power Spectrum}

The angular power spectrum of the SZ effect is defined in terms of the
multipole moments $y_{lm}$  of temperature fluctuations as
\begin{equation}
\left< y_{l m}^* y_{l' m'}\right> = C_l^\sz \delta_{l l'}\delta_{m m'}\,.
\end{equation}
$C_l^\sz$ is numerically equal to the flat-sky power spectrum in the flat sky limit.
The SZ power spectrum can be written as a redshift projection of the
pressure power spectrum
\begin{equation}
C_l^\sz = \int d\rad \frac{W^\sz(\rad)^2}{d_A^2} P_{\Pi\Pi}^\tot\left(\frac{l}{d_A},\rad\right),
\label{eqn:szpower}
\end{equation}
where $d_A$ is the angular diameter distance. At RJ part of the
frequency spectrum,  the SZ weight function
is
\begin{equation}
W^\sz(\rad) = -2 \frac{k_B \sigma_T \bar{n}_e}{a(\rad)^2 m_e c^2}
\end{equation}
where $\bar{n}_e$ is the mean electron density today. In deriving
Eq.~(\ref{eqn:szpower}),
we have used the Limber approximation \cite{Lim54} by setting
$k = l/d_A$ and flat-sky approximation. In previous studies (e.g.,
\cite{KomKit99} and references therein), the SZ power spectrum
due to massive halos have been calculated following projected $y$
parameter of individual halos. The two approaches are essentially 
same since the order in which the projection is taken does not matter,
except that our approach allows us to calculate intermediate 3d properties of
baryons, mainly pressure.

In Fig.~\ref{fig:szpower}(a), we show the  SZ power spectrum due to
baryons present in virialized halos compared with our previous
prediction for SZ effect using a biased power spectrum for pressure
fluctuations following \cite{PeaDod96} non-linear dark matter
power spectrum. As shown, most of the contributions to SZ power
spectrum comes from individual massive halos, while the halo-halo
correlations only contribute at a level of 10\% at large angular
scales. This is contrary to, say, the lensing convergence power
spectrum discussed in \cite{CooHu00b}, where most of the power
at large angular scales is due to the halo-halo correlations. The
difference can be understood by noting that the SZ effect is strongly
sensitive to the most massive halos due to $T \propto M^{2/3}$
dependence in temperature and to a lesser, but somewhat related, extent that its
weight function increases towards
low redshifts. The lensing weight function selectively probes
the large scale dark matter density power spectrum at comoving
distances half to that of background sources ($z \sim 0.2$ to 0.5 when
sources are at a redshift of 1), but has no extra dependence on mass.
We have also shown current upper limits on the temperature
fluctuations at arcminute scale angular scales where potentially the
physical properties of baryons can be studies. These limits come from
\cite{Holetal99} (BIMA) and \cite{Subetal00} (ATCA).

Also shown is the contribution to SZ effect from baryons present in
 overdensities $\lesssim 10$ (curve labeled GH). The SZ power spectrum
 here was calculated by replacing the pressure power spectrum in
 Eq.~(\ref{eqn:szpower}) with the unbiased
Jeans-scale smoothed dark matter power spectrum following
\cite{GneHui98} and assuming a mean temperature of 25 eV for
 these baryons. The mass fraction of baryons present in such small
 overdensities were taken from numerical simulations of
 \cite{CenOst99} and roughly follows $\sim 0.25(1+z)$, such that
 at a redshift of 3 and above all of the baryons are present  in such
 small overdensities. The power spectrum due to such baryons are
 roughly three orders of magnitude lower than the power spectrum
 predicted for SZ effect from baryons in virialized halos, but as
 shown in Fig.~\ref{fig:szpower}(b), this level is consistent with
what is predicted for SZ effect when halos  with mass greater than
 $10^{13}$ M$_{\sun}$ is not present in observed fields. 

As shown in Fig.~\ref{fig:szpower}(b), 
the lack of massive halos leads to a strong suppression of power, and
 halos with masses greater than $10^{15}$ M$_{\sun}$ are needed to
 obtain the full power spectrum. 
The lack of massive halos not only lead to a change in the power
spectrum at large angular scales, the lack of mases also affect the
contribution at small angular scales. Increasing the minimum
temperature of electrons from the values determined by virial theorem
to a minimum energy value of 0.75 keV significantly affects the change
resulting from the lack of massive halos. In fact, with a minimum
energy for baryons, the change is smaller when halos with masses less
than $10^{14}$ M$_{\sun}$ is considered. At the higher end of masses,
the minimum energy do not significantly affect the power spectrum; the
resulting change is less than 30\% compared to the power spectrum
with electron temperature based on the virial theorem. The variations
suggest several observational possibilities, including  the
determination of minimum electron temperature, ie. the energy related
to preheating if it exists, by calculating the power spectrum
with massive halos substratced in a wide-field SZ map such as the one
that will be eventually made with Planck.

For less area surveys, such as planned interferometric observations of
the wide-field SZ effect (e.g., the few square degree survey of
Carlstrom et al. \cite{Caretal96}),
the sample variance due to lack of massive halos in observed fields can be problematic in the
interpretation of the observed signal.  The problem arises from the
fact that massive halos which contribute to the SZ power spectrum are
rare and that observations in small fields will not contain such
adeqaute masses to provide the fully expected SZ signal. The 
dependance of SZ effect on massive halos is even problematic for
numerical simulations with limited box sizes. As pointed out by
\cite{Seletal00}, the measured power spectrum in their simulation
varies significantly based on the considred line of sight.

The dependance of signal on massive halos is also present in other
observables of large scale structure, such as weak gravitational lensing.
Compared to weak lensing surveys, studied in \cite{CooHu00b} and \cite{Cooetal00b},
the SZ effect depends more strongly on rare halos.  Most of these
halos are at low redshifts, thus, surveys which avoid regions
with known clusters will inherently also include an additional bias.
As an example, the contribution to 1-$\sigma$ detection of temperature anisotropies at arcminute scales
by \cite{Holetal99} due to SZ effect requires detailed knowledge on
the distribution of halo masses in the observed fields. For a
measurement of the SZ power spectrum, with a sample variance less than
20\%, requires observations of a field $\sim$ 1000 deg$^2$, while the
same can be achieved in an area of $\sim$ 100 deg$^2$ for lensing.
As discussed in  \cite{CooHu00b}, however,
the sample variance due to lack of massive and rare halos,
which dominate the SZ power, does not directly
 imply a systematic bias as long as one uses an approach similar to the one
suggested by carefully accounting for the sample variance that may be
 present from lack of massive halos.  Such an approach requires a
reliable model for the SZ effect and detailed numerical simulations
will be required for such a study.

These issues can be ignored for the upcoming wide field CMB experiments, such as Planck, where the
frequency coverage will allow the recovery of SZ effect over 65\% of
the sky not confused by galactic emissions,
thereby, providing an accurate measurement of its power
spectrum and higher order statistics (see, \cite{Cooetal00a}). Such a
wide-field SZ map is also highly desirable for several reasons
including the presence of adequate mass distribution of the universe
such that a fair sample is considered and the possibility to use such
a wide-field map for various cross-correlations purposes, such as
against Sloan galaxy distribution or a wide-field weak lensing survey
(see, \S~\ref{sec:sz-lens} and \ref{sec:sz-gal}).

\subsection{SZ Bispectrum \& Skewness}

The angular bispectrum of the SZ effect is defined as
\begin{equation}
\left< y_{l_1 m_1} y_{l_2 m_2} y_{l_3 m_3} \right>
= \wjm B_{l_1 l_2 l_3}^\sz \,.
\end{equation}
and can be written following Limber approximation as
\begin{eqnarray}
\bi^\sz &=& \sqrt{(2l_1+1)(2l_2+1)(2l_3+1)  \over 4\pi} \wj
        \nonumber\\
&&\times \left[ \int dr {[W^\sz(r)]^3 \over \da^4}  B_\Pi^\tot\left({l_1 \over
        \da},{l_2 \over \da},{l_3\over \da};r\right)\right] \, .
\label{eqn:szbispectrum}
\end{eqnarray}
The more familiar flat-sky bispectrum is simply the expression in
brackets \cite{Hu00}.
The basic properties of Wigner-3$j$ symbol
introduced above can be found in \cite{Cooetal00a}.

Similar to the density field bispectrum,
we define
\begin{equation}
\Delta^2_{{\rm eq}l} = \frac{l^2}{2 \pi}
\sqrt{|B^\sz_{l l l}|} \, ,
\end{equation}
involving equilateral triangles in $l$-space. The absolute value of
$B^\sz$ is considered in above since $B^\sz \propto g(x)^3$, which is a 
negative quantity at RJ part of the frequency spectrum with $g(x)=-2$.

In Fig.~\ref{fig:szbispectrum}(a), we show individual contributions
$\Delta^2_{{\rm eq}l}$ with a maximum mass of $10^{16}$ M$_{\sun}$.
As shown, the main contribution to bispectrum comes from individual
halo term, while other terms involving correlations between halos
contribute $\lesssim 10$\%. In Fig.~\ref{fig:szbispectrum}(b), we
show bispectrum as a
function of maximum mass used in the calculation. Here, we have shown
the total contribution to the bispectrum in solid lines while the
total bispectrum in the presence of a minimum temperature of 0.75 keV
is shown with a dot-dashed line. The bispectrum, as discussed in
\S~\ref{sec:results}, is strongly sensitive to the single halo term due
to additional mass weighing. Almost all of the contributions to the SZ
bispectrum comes from the single halo term. The same dependence on
mass massive rare halos decreases the effect of a temperature increase
when maximum mass in $\gtrsim 10^{15}$ M$_{\sun}$. This is in contrast
to the power spectrum, where differences are still present with an
increase in temperature from virial to a minimum of 0.75 keV.

The measurement of full bispectrum in million to billion pixel data of
a wide-field SZ map as the one that'll be produced with Planck
can in general be difficult. In fact, there is no algorithm yet to
measure the full bispectrum in such wide-field data in a reasonable
time and computational requirements. Given such a possibility, it is
interesting to consider a collapsed measurement of the bispectrum;
real space statistics such as the third moment and skewness allow this
possibility. In fact, the skewness has now been measured for the COBE data by
\cite{Conetal00}, while the bispectrum measurements have only been
limited to specific configurations of the bispectrum such as
equilateral triangles in $l$-space \cite{Feretal98}.
The skewness allows an easily measurable 
aspect of the bispectrum and will be porbably be one of the first
measurements of non-Gaussianity in a wide-field SZ map. The skewness
can be calculated using the second, $\left< y^2(\sigma) \right>$, and
third, $\left< y^3(\sigma) \right>$, moments of the SZ effect:
\begin{equation}
S_3(\sigma) =
\frac{\left<y^3(\sigma)\right>}{\left<y^2(\sigma)\right>^2}
\, ,
\end{equation}
where the two moments are
\begin{eqnarray}
\left<y^3(\sigma) \right> &=&
                {1 \over 4\pi} \sum_{l_1 l_2 l_3}
                \sqrt{(2l_1+1)(2l_2+1)(2l_3+1) \over 4\pi} \nonumber\\
                &&\times \wj \bi^\sz
                W_{l_1}(\sigma)W_{l_2}(\sigma)W_{l_3}(\sigma)
                \,. \nonumber \\
\end{eqnarray}
and
\begin{equation}
\left< y^2(\sigma) \right> =
{1 \over 4\pi} \sum_l (2l+1) C_l^\sz W_l^2(\sigma)\,.
\label{eqn:secondmom}
\end{equation}

In Fig.~\ref{fig:szskewness}(a), we show the absolute value of
skewness, $|S_3(\sigma)|$, as a function of smoothing scale $\sigma$
when the maximum mass included in the calculation ranges from
$10^{16}$ to $10^{13}$ M$_{\sun}$ (from top to bottom). 
The absolute value of
skewness is considered since $S_3 \propto g(x)^{-1}$, which is a 
negative quantity at RJ part of the frequency spectrum with $g(x)=-2$.
As shown, the SZ skewness is heavily dependent on the presence of most massive and rare
halos. The introduction of a minimum energy of 0.75 keV leads to
decrease in skewness, which results from the fact that the power
spectrum is more affected than the bispectrum by such an increase.

Given the dependence on rare and most massive halos, the SZ effect
is considerably non-Gaussian. As discussed in \cite{AghFor99},
the non-Gaussianity can be used as a useful tool for the
identification and separation of most rare and massive halos from the 
wide-field CMB data.  An optimised algorithm that utilised both the
non-Gaussianity of SZ effect and its frequency dependance will be
useful for constructing a catalog of SZ clusters in upcoming
wide-field data. With massive clusters separated out, the remaining contribution
to SZ effect will be from halos of mass $\lesssim 10^{14}$ M$_{\sun}$,
such as galaxy groups. In \cite{Cooetal00a}, we considered
contribution from such small halos as the one due to large scale
structure. The extent to which such small halos contribute to the SZ
power spectrum and higher order statistics clearly depend on the role of
additional energy in baryons. Therefore, any measurement of the power
spectrum with known massive clusters removed, can be in return used as
a probe of physical properties related to baryons, mainly the extent
to which prehating affects the electron temperature of low mass halos.

\begin{figure*}[t]
\centerline{\psfig{file=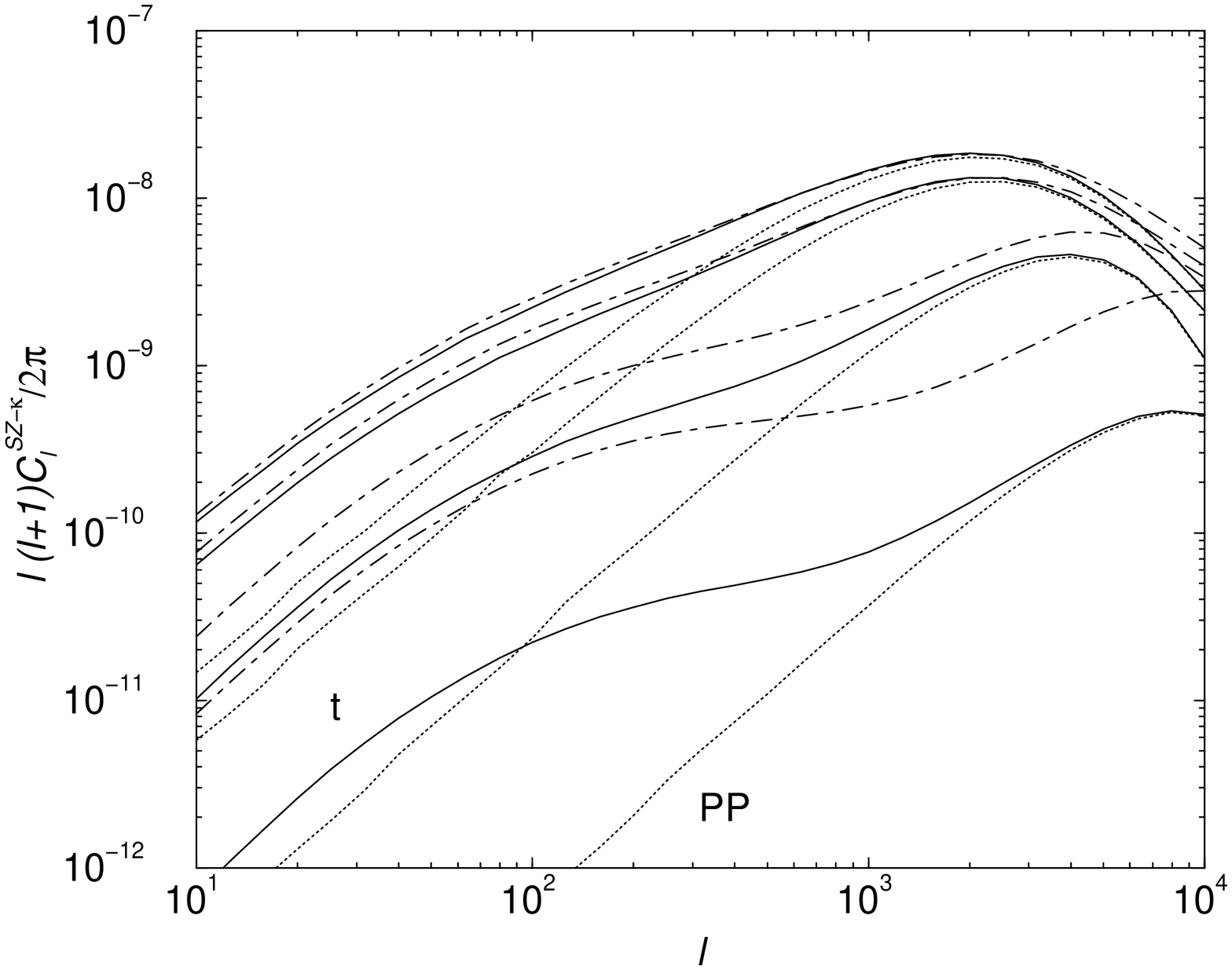,width=3.9in,angle=-90}
\psfig{file=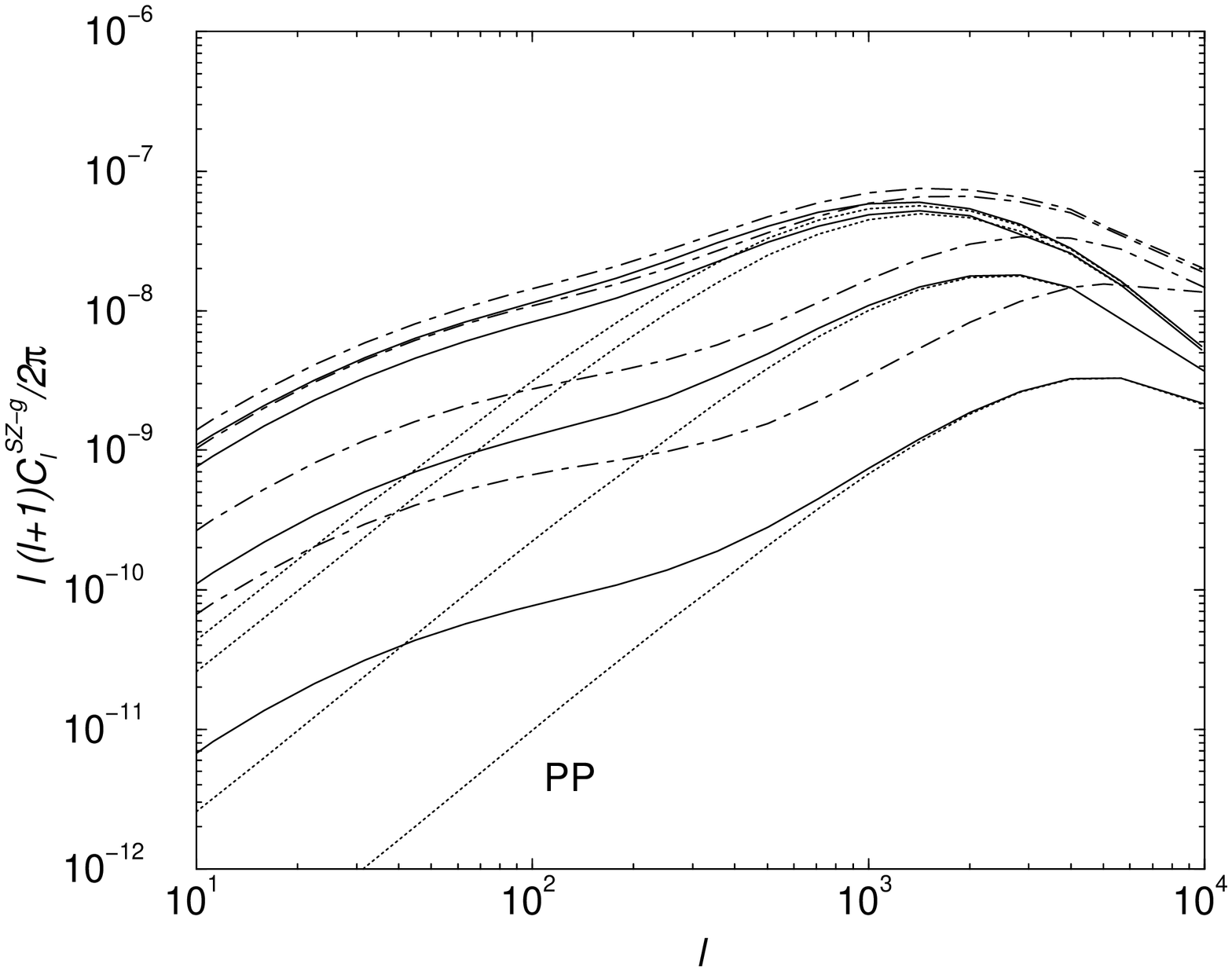,width=3.9in,angle=-90}}
\caption{SZ-weak lensing cross-correlation (a) and 
SZ-galaxy cross-correlation (b) power spectra as a function of maximum
mass, with mass cut off following Fig.~1. The additional minimum
energy, shown with a dot-dashed line, leads to an increase in the
correlated power.
}
\label{fig:szlens}
\end{figure*}

\begin{figure}[t]
\centerline{\psfig{file=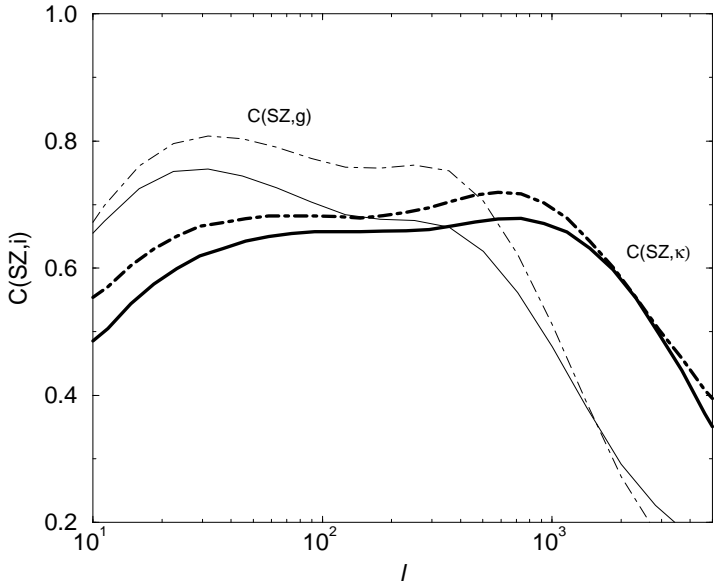,width=3.7in,angle=-90}}
\caption{
The cross-correlation coefficient for SZ-weak lensing
(thick lines) and SZ-galaxy (thin lines) 
with (solid line) and without additional energy (dashed line). 
In general, when the maximum mass is greater than $10^{14}$ M$_{\sun}$,
SZ and galaxy correlate less than SZ-weak lensing at small
angular scales, while the opposite is present at large angular scales.
Introduction of additional energy increases the correlation by not more than few
percent; such an increase is insignificant to be determined
observationally using the SZ-lensing and SZ-galaxy cross-correlations,
however, SZ power spectrum and skewness allow a determination as these
values are significantly affected.}
\label{fig:szcorr}
\end{figure}

Instead of individual non-Gaussian statistics as skewness, one can 
construct the probability distribution function (pdf) using a SZ map 
smoothed on some scale $\sigma$. The use of pdf as a probe of
cosmology was first suggested by \cite{Jaivan99} for weak gravitational
lensing convergence. The same technique can be easily extended to SZ.
Using the Edgeworth expansion to capture small deviations from
Gaussianity, one can write the pdf of SZ effect to second order as
\begin{eqnarray}
p(y) &=& \frac{1}{\sqrt{2 \pi \left<y^2(\sigma)\right>}}
e^{-y(\sigma)^2/2\left< y^2(\sigma)\right>}  \nonumber \\
&\times& \left[ 1+ \frac{1}{6} S_3(\sigma) \sqrt{\left< y^(\sigma)
\right>} H_3\left(\frac{y(\sigma)}{\sqrt{\left< y^2(\sigma) \right>}} \right)\right] \, ,
\end{eqnarray}
where $H_3(x) = x^3 -3x$ is the third order Hermite polynomial (see,
\cite{Jusetal95} for details).

In Fig.~\ref{fig:szskewness}(b), we show the pdf of SZ effect at 12$'$
as a function  of maximum mass used in the calculation. As shown, the
greatest departure from non-Gaussianity occur when the maximum mass
of halos are greater than $10^{14}$ M$_{\sun}$. Given that we have
only constructed the pdf using terms only out to skewness, the
presented
pdfs should only be consider as approximate; With increasing
non-Gaussianity behavior, the approximated pdfs are likely to depart
from true distributions especially in the tails. 
Observationally, the pdf can be constructed easily by considering a
histogram of the pixel temperature values of the SZ map.
Though such a construction sounds straightforward, there is likely to
be complications on the interpretation of such a pdf in the presence
of instrumental noise and other foregrounds. Techniques which do no
directly make a wide-field map, especially interferometric
observations, will again require special techniques to construct the pdf.
Therefore, the extent to which the full pdf, or such an histogram, 
can be used a probe of cosmology and the accuracy to which
pdfs can be constructed  from upcoming wide-field CMB anisotropy data,
such as Planck and planned interferometric surveys (Carlstrom, private
communication), need to be investigated in detail. We leave these
issues for further study.

\subsection{SZ-Weak Lensing Cross-correlation}
\label{sec:sz-lens}

Similar to the SZ power spectrum, the angular power spectrum of
weak lensing convergence can be defined in
terms of the multipole moments $\kappa_{l m}$ as
\begin{equation}
\left< \kappa_{l m}^* \kappa_{l' m'}\right> = C_l^\kappa \delta_{l l'}
\delta_{m m'}\, \, ,
\end{equation}
and can be written in terms of the dark matter power spectrum by
\cite{Kai92,Kai98}
\begin{equation}
C^\kappa_l = \int d\rad \frac{W^\kappa(\rad)^2}{d_A^2}
P_\delta^\tot\left(\frac{l}{d_A};\rad\right) \, ,
\label{eqn:lenspower}
\end{equation}
When all background sources are at a distance of
$\rad_s$, the lensing weight function becomes
\begin{equation}
W^\kappa(\rad) = \frac{3}{2} \Omega_m \frac{H_0^2}{c^2 a} \frac{
d_A(\rad) d_A(\rad_s -\rad)}{d_A(\rad_s)} \, .
\end{equation}
The detail properties of lensing statistics, under the dark matter
halo approach, is discussed in \cite{Cooetal00b} and
\cite{CooHu00b}.

The cross-correlation between the SZ effect and weak gravitational
lensing can be similarly defined in terms of the individual multipole
moments as
\begin{equation}
\left< \kappa_{l m}^* y_{l' m'}\right> = C_l^{\sz-\kappa} \delta_{l l'}
\delta_{m m'}\, \, .
\end{equation}
This is now related to the dark matter-pressure power spectrum by 
\begin{equation}
C_l^{\sz-\kappa} = \int d\rad \frac{W^\sz(\rad)W^\kappa(\rad)}{d_A^2}
P_{\Pi\delta}^\tot\left(\frac{l}{d_A};\rad\right) \, .
\label{eqn:szlenspower}
\end{equation}
Finally the cross-correlation coefficient between SZ and weak lensing
is
\begin{equation}
C(\sz,\kappa)_l  = \frac{C^{\sz-\kappa}_l}{\sqrt{C^\sz_l C^\kappa_l}}
\end{equation}

\subsection{SZ-Galaxy Cross-correlation}
\label{sec:sz-gal}

Similar to the SZ-weak lensing cross-correlation, one can study the
cross-correlation
between the galaxy distribution, which traces the large scale
structure, and the SZ effect.
The power spectrum of galaxy distribution can be defined
terms of the multipole moments $g_{l m}$ as
\begin{equation}
\left< g_{l m}^* g_{l' m'}\right> = C_l^g \delta_{l l'}
\delta_{m m'}\, \, ,
\end{equation}
and can be written as a projection of the dark matter power spectrum
\begin{equation}
C^g_l = \int d\rad \frac{W^g(\rad)^2}{d_A^2}
P_\delta^\tot\left(\frac{l}{d_A};\rad\right) \, .
\label{eqn:galpower}
\end{equation}
Here, $C_l^g$  should be understood as the 2d Fourier transform of the
galaxy correlation function, generally referred to as $w(\theta)$ in
the literature.
The weight function for galaxy projection involves the redshift distribution
\begin{equation}
W^\kappa(\rad) = \frac{dN}{d\rad} \,
\end{equation}
normalized such that $\int d\rad (dN/d\rad) =1$.

As before. the cross-correlation between the SZ effect and galaxy distribution 
can be similarly defined in terms of the individual multipole moments as
\begin{equation}
\left< g_{l m}^* y_{l' m'}\right> = C_l^{\sz-g} \delta_{l l'}
\delta_{m m'}\, \, .
\end{equation}
This is now related to the dark matter-pressure power spectrum by 
\begin{equation}
C_l^{\sz-g} = \int d\rad \frac{W^\sz(\rad)W^g(\rad)}{d_A^2}
P_{\Pi\delta}^\tot\left(\frac{l}{d_A};\rad\right) \, .
\label{eqn:szgalpower}
\end{equation}
Finally the cross-correlation coefficient between SZ and weak lensing
is
\begin{equation}
C(\sz,g)_l  = \frac{C^{\sz-\kappa}_l}{\sqrt{C^\sz_l C^g_l}}
\end{equation}

In Fig.~\ref{fig:szlens}, we show the SZ-weak lensing  and SZ-galaxy
cross-correlation power spectra
as a function of maximum mass used in the calculation, while the
correlation coefficients  are shown in Fig.~\ref{fig:szcorr}. 
In order to describe the galaxy distribution, we have considered a
survey at low redshifts, similar to the Sloan Digital Sky Survey 
(SDSS)\footnote{http://www.sdss.org}. Such a low redshift tracer is
desirable since contributions to SZ effect primarily comes from large
scale structures at redshifts $< 1$. Here, we have assumed the
redshift distribution of Sloan galaxies follow $dN/dz \propto z^2
\exp[(-z/z_c)^{3/2}]$ with a mean redshift $z_m$, of 0.2 ($z_c \sim 1.412 z_m$).
For SZ-weak lensing, the cross-correlation is such that
SZ and lensing traces each other out to angular scales of $\sim$ 1000
when most massive and rarest halos are involved with a decrease in
cross-correlation between the two at small angular scales. 
For SZ-galaxy cross-correlation, there is additional correlation at
large angular scales, when compared to lensing, while the correlation
is suppressed at small angular scales. The decrease at small angular
scales is due to the fact that small halos that contribute to SZ and
lensing do not necessarily contribute to the galaxy power spectrum.
With an increase in additional energy for halos, the cross-correlation
increases by few percent, however, this small increase unlikely to be
determined accurately through observations. 
The correlation is sensitive to the redshift distribution of galaxies,
which depends mostly on the selection criteria imposed by observations.
The selection function of weak lensing can be considered well
understood, however, a straight forward interpretation of any observed
SZ-galaxy cross-correlation will require a clear understanding of
observable related to galaxy distribution.

The SZ-SZ, lensing-lensing and SZ-lensing power spectra dependent
different on the bias and correlation parameters. Since the bias
and correlation are scale and redshift dependent, the measurement
of these power spectra, which are projected along the redshift
distributions, do not allow a direct probe of these quantities.
A useful approach would be to consider the inversion of these
power spectra in redshift bins by considering the measured
lensing-lensing and SZ-lensing power spectra as a function of redshift. Note
that the SZ-SZ power spectrum 
cannot be easily separated in redshift space as we do
not have the ability to separate individual redshift contributions,
unlike say in lensing, where one can use the redshifts of background
sources to construct convergence as a function of redshift.
With adequate signal-to-noise from wide-field surveys, it is likely
that such an approach will allow studies to be carried out on the
extent to which temperature weight baryons trace the dark matter and
their correlation properties. 

Similarly, as studied in \cite{PeiSpe00}, the
cross-correlation of SZ against galaxy data, as a function of
redshift, is expected to provide information on the
properties of clustering of galaxies with respect to the temperature
weighted baryon  field represented by SZ effect. Such  a
cross-correlation will help understand the extend to which hot/warm
gas is present in the outskirts of individual galaxies. 
With individual sets of SZ, weak lensing and galaxy  maps,
it is likely that a tremendous amount of information on physical
properties associated with dark matter, baryon and galaxy distribution
will be obtained through both a comparison of individual power spectra
and higher order moments and cross-correlations and higher order
moments associated with such cross-correlations. We hope to study some
of these possibilities in detail in future studies.

\section{Summary \& Conclusions}
\label{sec:conclusions}

Using an extension of the dark matter halo approach, we have presented
an efficient method to calculate the large scale structure pressure
power spectrum and its high order moments, such as bispectrum. 
We have divided the contribution to large scale pressure power
spectrum based on the overdensities in which contributing baryons are
present with (1) baryons
present in virialized halos with overdensities greater than $\sim$ 200
and in hydrostatic equilibrium with the density field of such halos,
(2) photoionized baryons in overdensities less than $\sim$ 10 and 
which trace the Jeans-scale smoothed dark matter  density field, and
(3) baryons present in the mid overdensity regime which are likely to be
undergoing collapse and shock heating. 

Our approach allows us to calculate not only 2d statistics such as the
projected pressure power spectrum, or the SZ effect, which will be
observed, but also the 3d statistics that can be directly compared to
predictions based on numerical simulations. We have performed such a
comparison to recently published numerical simulations by
\cite{Refetal99} and found good agreement between our analytical
calculations and their simulations. The current simulations are
limited to a handful of realizations and limited dynamical range and
resoltion. With improving resolution and accuracy, analytical models
such as the one presented here will be tested in detail against
numerical calculations. Analytical calculations, aided by numerical
simulations, will eventually allow detailed studies of large scale
baryon distribution using observations such as the wide-field SZ effect.

The projected pressure power spectrum along the line of sight,
provides a direct calculation of the large scale structure SZ effect
and its higher order correlations.
In the absence of massive and rare halos, we  have suggested that
baryons present in small overdensities provide a lower limit to any
contribution to SZ effect. The extent to which baryons present in
overdensities between 10 and 200 contribute to the correlations in large scale
pressure and, from it, the SZ effect, requires additional studies,
preferably with numerical simulations. Presently, our
Understanding the role of preheating and its effect of
baryons will also be another challenge as the SZ observations will
clearly depend on such additional energy contributions to large scale
baryon distribution. We have suggested the possibility of using SZ
power spectrum and higher order correlations, such as the SZ skewness,
as a probe of preaheating. Such a study will require a wide-field SZ
map and this task will be completed with Planck observations. The unique
frequency dependance of the SZ effect, together with its non-Gaussian
behavior, will allow the construction of a reliable SZ cluster catalog
with will aid in cosmologicalo studies of structure formation.

Our approach allows one to study possible systematic effects that may
be present in upcoming SZ observations of small area fields due to the presence
or absence of rare massive halos in such fields that will be
observed. We have shown that the SZ effect as well as its
non-Gaussian properties are mainly due to
the most massive and rarest virialized halos in the universe. 
The lack of massive halos in observed SZ fields can introduce
a systematic bias in the power spectrum, but the sample variance
introduced by the lack of such masses, can be easily corrected based
on the prior knowledge on mass distribution of observed fields.
Due to additional mass dependence through temperature, the 
effect of mass is such that the SZ effect is more dependent
on the rare halos than weak gravitational lensing convergence.
The same SZ halos also contribute to lensing convergence and the
cross-correlation between SZ and lensing can be used a probe of
clustering properties between density and temperature weighted baryon
fields.  Given the great potential to study baryon distribution using
SZ, various issues suggested here involving such correlations 
merit further study.

\acknowledgments
The author greatly thanks Wayne Hu for useful discussions and helpful
suggestions that led to the calculations presented in this paper.
We also acknowledge useful discussions with  Gil Holder, Lloyd Knox and Joe
Mohr and thank Alexandre Refregier, Ue-Li Pen and their collaborators
 for providing results from numerical simulation presented in
\cite{Refetal99}.

\end{document}